\pdfoutput=1
\documentclass[aps,prl,twocolumn,superscriptaddress,showpacs]{revtex4-1}

\usepackage{graphicx}%include figure files
\usepackage{afterpage}
\def\etal{{\em{et al.}}}

\begin{document}

%\preprint{}

\title{Competing magnetic states, disorder, and the magnetic character
of Fe$_3$Ga$_4$}



\author{J. H. Mendez}
%\email[]{ditusa@phys.lsu.edu}
%\thanks{}
\affiliation{Department of Physics \& Astronomy, Louisiana State
University, Baton Rouge, LA 70803, USA}

\author{C. E. Ekuma}
%\thanks{}
\affiliation{Department of Physics \& Astronomy, Louisiana State
University, Baton Rouge, LA 70803, USA} 
\affiliation{Center for
Computation and Technology, Louisiana State University, Baton Rouge,
LA 70803, USA}

\author{Y. Wu}
%\thanks{}
\affiliation{Department of Physics \& Astronomy, Louisiana State
University, Baton Rouge, LA 70803, USA}

\author{B. Fulfer}
%\thanks{}
\affiliation{Department of Chemistry, Louisiana State
University, Baton Rouge, LA 70803, USA}

\author{J. C. Prestigiacomo}
%\thanks{}
\affiliation{Department of Physics \& Astronomy, Louisiana State
University, Baton Rouge, LA 70803, USA}

\author{W. A. Shelton}
%\thanks{}
\affiliation{Center for Computation and Technology, Louisiana State
University, Baton Rouge, LA 70803, USA} 
\affiliation{Department of
Chemical Engineering, Louisiana State University, Baton Rouge, LA
70803, USA}

\author{M. Jarrell}
%\thanks{}
\affiliation{Department of Physics \& Astronomy, Louisiana State
University, Baton Rouge, LA 70803, USA} 
\affiliation{Center for
Computation and Technology, Louisiana State University, Baton Rouge,
LA 70803, USA}

\author{J. Moreno}
%\thanks{}
\affiliation{Department of Physics \& Astronomy, Louisiana State
University, Baton Rouge, LA 70803, USA} 
\affiliation{Center for
Computation and Technology, Louisiana State University, Baton Rouge,
LA 70803, USA}

\author{D. P. Young}
%\thanks{}
\affiliation{Department of Physics \& Astronomy, Louisiana State
University, Baton Rouge, LA 70803, USA}

\author{P. W. Adams}
%\thanks{}
\affiliation{Department of Physics \& Astronomy, Louisiana State
University, Baton Rouge, LA 70803, USA}

\author{A. B. Karki}
%\thanks{}
\affiliation{Department of Physics \& Astronomy, Louisiana State
University, Baton Rouge, LA 70803, USA}

\author{R. Jin}
%\thanks{}
\affiliation{Department of Physics \& Astronomy, Louisiana State
University, Baton Rouge, LA 70803, USA}

\author{Julia Y. Chan}
%\thanks{}
\affiliation{Department of Chemistry, University of Texas at Dallas,
Richardson, TX 75080, USA}

\author{J. F. DiTusa}
\email[]{ditusa@phys.lsu.edu}
%\thanks{}
\affiliation{Department of Physics \& Astronomy, Louisiana State
University, Baton Rouge, LA 70803, USA}


\date{\today}

\begin{abstract}
The physical properties of metamagnetic Fe$_3$Ga$_4$ single crystals
are investigated to explore the sensitivity of the magnetic states to
temperature, magnetic field, and sample history. The data reveal a
moderate anisotropy in the magnetization and the metamagnetic critical
field along with features in the specific heat at the magnetic
transitions $T_1=68$ K and $T_2=360$ K. Both $T_1$ and $T_2$ are found
to be sensitive to the annealing conditions of the crystals suggesting
that disorder affects the competition between the ferromagnetic (FM)
and antiferromagnetic (AFM) states. Resistivity measurements reveal
metallic transport with a sharp anomaly associated with the transition
at $T_2$. The Hall effect is dominated by the anomalous contribution
which rivals that of magnetic semiconductors in magnitude ($-5 \mu
\Omega$ cm at 2 T and 350 K) and undergoes a change of sign upon
cooling into the low temperature FM state. The temperature and field
dependence of the Hall effect indicate that the magnetism is likely to
be highly itinerant in character and that a significant change in the
electronic structure accompanies the magnetic transitions. We observe
a contribution from the topological Hall effect in the AFM phase
suggesting a non-coplanar contribution to the magnetism. Electronic
structure calculations predict an AFM ground state with a wavevector
parallel to the crystallographic $c$-axis preferred over the
experimentally measured FM state by $\approx$ 50 meV per unit
cell. However, supercell calculations with a small density of
Fe-antisite defects introduced tend to stabilize the FM over the AFM
state indicating that antisite defects may be the cause of the
sensitivity to sample synthesis conditions.
\end{abstract}

%pick put some pac numbers later
%\keywords{}

\maketitle

%\section{}
%\subsection{}
%\subsubsection{}
\section{Introduction}
Metallic antiferromagnets have received renewed attention over the
past few years, in part, because the iron pnictide and chalcogenide
families of superconductors are derived via chemical substitutions
into metallic antiferromagnetic parent
compounds\cite{hosono,paglione,stewart}. The character of the magnetic
state in these materials, spin density wave (SDW) or more conventional
local antiferromagnetism, has been explored and argued over as it has
implications for the superconducting pairing
mechanism\cite{eremin}. Further fueling this interest is the discovery
of the coexistence of itinerant ferromagnetism and local
antiferromagnetism in the related material
Ba$_{1-x}$K$_x$Mn$_2$As$_2$\cite{pandey}. More recently and, perhaps
more unexpectedly, the titanium-based pnictide oxide
Ba$_{1-x}$Na$_x$Ti$_2$Sb$_2$O was discovered to have density wave
states, spin (SDW) and/or charge, that coexist with a low temperature
superconducting state\cite{lorenz}. This activity has built on a long
history of exploration of antiferromagnets related to the cuprate
superconductors as the interesting magnetic properties of these
compounds are thought to be of central importance to their
unconventional superconducting states\cite{birgeneau}. The difficulty
in separating out the important aspects of these complex materials has
driven explorations of simpler antiferromagnetic metals such as
elemental chromium\cite{kummamuru}, a prototypical spin density wave
material, and GdSi\cite{feng}, a somewhat more complex system that has
both itinerant and local magnetic moments participating in the
magnetic ordering.

Here, we explore the properties of the lightly investigated Fe-based
binary Fe$_3$Ga$_4$. This compound is both metallic and magnetic and
there is a likely interdependence of local and itinerant magnetic
moments that determine its magnetic state. In this material the close
competition between antiferromagnetic (AFM) and ferromagnetic (FM)
states is evidenced by several transitions between them as well as the
sharp transition between the AFM phase and a field polarized
paramagnetic (PM) phase that occurs with the application of magnetic
field\cite{wagini1,wagini2}. Our data and simulations indicate that
this competition between such obviously different magnetic states
results from both its complex crystal structure, with its four unique
crystallographic Fe sites each with a somewhat different magnetic
moment and a large number of possible nearest-neighbor and
next-nearest-neighbor interactions\cite{philippe1,philippe2}, and the
properties of the itinerant charge carriers. One indication of this
coupling between the itinerant and more local moments is the
temperature dependence of the metamagnetic field, $H_{mm}$, the field
necessary to drive the transition from AFM to PM with a FM alignment
of the field induced moments, which has the unusual feature that it
increases with temperature. Although the phenomenology of itinerant
metamagnetism was worked out decades
ago\cite{moriyamm,isoda,stryjewski}, in practice there are several
different mechanisms that can cause abrupt transitions between AFM and
field polarized PM states with field.  For example, in
(Hf$_{1-x}$Ta$_x$)Fe$_2$ the symmetry of the crystal structure creates
a magnetic frustration at one of the two crystallographically distinct
Fe-sites. The magnetic moment at this site can be controlled via
doping\cite{nishihara}, and a larger magnetic moment at higher doping
favors ferromagnetism. In CoMnSi, the field polarized PM-to-AFM
instability appears to be closely related to the Mn-Mn separation
within the orthorhombic crystal structure so that thermal expansion or
chemical substitution causes an abrupt phase
change\cite{barcza,staunton}. These materials are of interest for
possible technological relevance as well since the closely competing
magnetic orderings could allow applications as magneto-caloric
elements\cite{gschneidner,nishihara,wada,rechenberg,barcza,staunton,sharma}.

%The crystal structure of magnetic materials plays an important role in
%determining the energy scales for magnetic interactions, the magnetic
%ground state structure, and their critical temperatures.  For example,
%in insulating systems with antiferromagnetic interactions and
%geometric frustration, such as in Kagome honeycomb lattice, or spin
%ice materials\cite{ramirez,karundasa,zvyagin}, the local magnetic
%moment site symmetry can delay ordering to temperatures much below the
%interaction scales and result in a significant magnetic entropy
%surviving to very low temperature, $T$. There are also systems where
%subtle changes in crystal structure with temperature or pressure
%result in very different magnetic states. Such competing magnetic
%orders often result in sensitivities to environmental parameters such
%as temperature, $T$, magnetic field, $H$, and pressure, $P$, so that
%sharp phase transitions among them can often be
%accessed\cite{stryjewski}. This sensitivity can be useful either as
%sensors or, in cases where the magneto-elastic coupling is strong, as
%magneto-caloric elements\cite{gschneidner}. Recently, several
%materials that display abrupt transitions from an antiferromagnetic
%(AFM) to a ferromagnetic (FM) state upon cooling or exposure to
%magnetic fields, commonly referred to as metamagnetic, have been
%considered for magneto-caloric applications. Example of such materials
%include (Hf$_{1-x}$Ta$_x$)Fe$_2$\cite{nishihara,wada,rechenberg},
%CoMnSi\cite{barcza,staunton}, and NiMn$X$ ($X$=In, Sn, and
%Sb)\cite{sharma}.

The magnetic properties of Fe$_3$Ga$_4$ have been previously
characterized by susceptibility, $\chi$, and magnetization, $M$,
measurements on polycrystalline samples which revealed several
magnetic
transitions\cite{wagini1,kawamiya1,kawamiya2,duijn,duijnthesis}
between FM, AFM, and field polarized PM states. The ground state is FM
with a transition to an AFM-like state near $T_1=68$ K. This is
accompanied by an unusual metamagnetism whereby $H_{mm}$ increases
dramatically with $T$ up to $T_2\sim 360$ K\cite{kawamiya1}. The
reduction of $H_{mm}$ with cooling indicates a continuous decrease in
the energy difference between the FM and AFM states until a
first-order transition at $T_1$ where the FM state emerges as the
ground state.  Above 400 K $\chi$ is reduced\cite{kawamiya1} and the
M\"{o}ssbauer spectra evolves into a single broad line\cite{kobeissi}
so that a critical temperature for magnetic ordering was identified at
$392$ K. M\"{o}ssbauer experiments have also established a different
magnetic moment for each of the four unique crystallographic
Fe-sites\cite{kawamiya1,kobeissi} increasing the complexity of this
binary system. Duijn \etal~\cite{duijn,duijnthesis} explored the
specific heat and thermal expansion of Fe$_3$Ga$_4$ finding only a
small anomaly in the thermal expansion at $T_1$. The ability to grow
polycrystalline grains of this material on GaAs substrates has led to
the discovery of photomagnetic effects where an illumination enhanced
magnetization was demonstrated\cite{munekata}. This effect is most
likely caused by simple heating through $T_2$, although the existence
of a photon-mode photo-enhanced magnetization has been
suggested\cite{jamil}.  Despite all of this interest, there have been
almost no explorations of Fe$_3$Ga$_4$ in single crystalline
form\cite{wagini1,wagini2} and the identity of the magnetic states has
not been established as neutron scattering experiments on powders were
inconclusive\cite{duijnthesis}.

We report on the magnetic, thermodynamic, and charge transport
properties in single crystals of Fe$_3$Ga$_4$ establishing a moderate
anisotropy of the magnetic properties and the magnetic phase diagram
for two orientations of an external magnetic field. We have carefully
measured the specific heat of these crystals identifying the magnetic
contributions at $T_1$ as well as the contribution above room
temperature that grows near $T_2$. In addition, we have measured the
resistivity, $\rho$, magnetoresistance (MR), and Hall effect of our
crystals.  The $\rho$ is metallic and marked by an abrupt change at
$T_2$ while the MR displays sharp changes at $H_{mm}$ both of which
broadly reproduce the main findings of measurements on the
polycrystalline samples\cite{kawamiya1,duijnthesis}. The $T$ and $H$
dependence of $\rho$ hint at a close interdependence of the charge
carriers and the ordered magnetic moments such as would occur in a SDW
material. Our Hall measurements reveal a large anomalous Hall effect
reaching -5 $\mu \Omega$cm in a 2 T field above room temperature, a
value more representative of a magnetic semiconductor\cite{manyala2},
and an ordinary Hall contribution consistent with a carrier
concentration of 1 carrier per formula unit. Both the anomalous and
ordinary terms show dramatic changes upon warming through $T_1$
including a change of sign for the anomalous Hall effect from positive
to negative, whilst the ordinary term remains positive. In addition,
we observe a significant topological Hall effect, $\rho_{THE}$,
indicating the possibility of a non-coplanar magnetic moment for
finite fields at $T_1 < T < T_2$\cite{ye,nagaosa}.

To gain further insight into the causes of the competition between the
magnetic phases and to help identify the magnetic order of the
intermediate phase, between $T_1$ and $T_2$, we have performed
extensive electronic structure calculations. These calculations predict
an AFM ground state with a wavevector along the crystallographic
$c$-axis in contrast to the experimentally observed FM state.
Furthermore, we show through supercell calculations performed with a
small density of Fe atoms replacing Ga to mimic antisite disorder that
the FM state can be stabilized. This suggests that disorder plays an
important role in this material, and perhaps in other itinerant
antiferromagnets, and can be used to effectively manipulate $T_1$ in
agreement with the observation of a sensitivity of this transition to
synthesis conditions.

\section{Experimental Details}
\begin{figure}[htb]
  \includegraphics[angle=0,width=3.2in,bb=70 15 630
  505,clip]{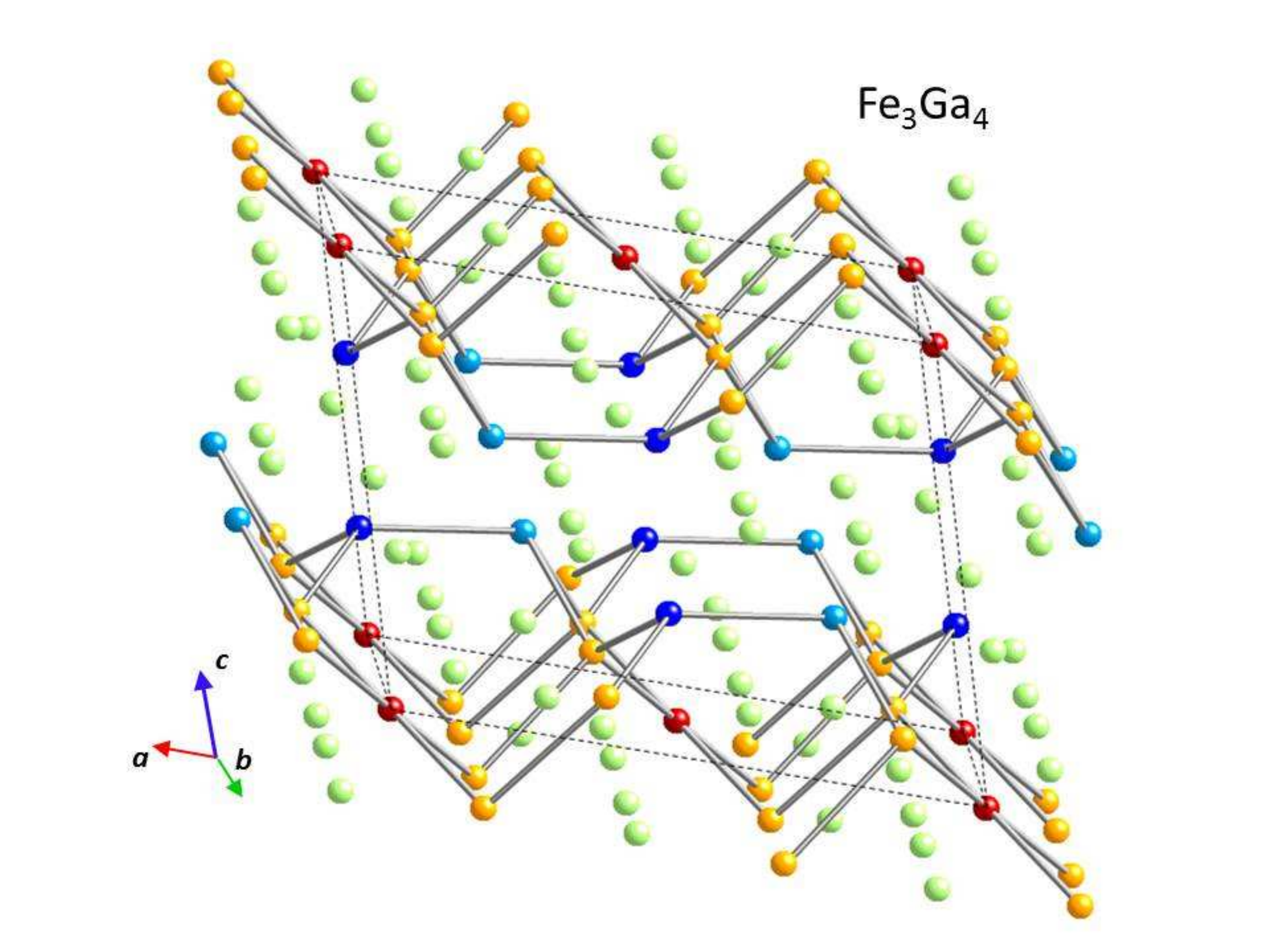}%
  \caption{\label{fig:struct} Fe$_3$Ga$_4$ crystal structure. The
 $C2/m$ base-centered monoclinic structure demonstrating the four
 unique Fe positions in the unit cell with Fe1 shown in red, Fe2 in
 dark blue, Fe3 in light blue, and Fe4 in orange. Ga atoms are shown
 in green. There are four unique Ga sites in this crystal structure
 that are not differentiated in the figure. The shortest Fe-Fe bonds
 (less than the 2.96 \AA between Fe2 sites) are highlighted.  }
\end{figure}

Single crystals of Fe$_3$Ga$_4$ were grown from high purity starting
materials by standard iodine vapor transport techniques at 750
$^\circ$C for 14 days\cite{philippe1,philippe2}.  These crystals are
black and shiny, and are roughly 1 mm by 1 mm by 2 mm thin brittle
plates.  All of the data presented in this paper were produced from
crystals grown via iodine vapor transport methods. We have also
employed optical furnace methods for synthesizing larger crystals and
the main results of the structural and magnetic measurements were
reproduced on these samples. Powder and single crystal X-ray
diffraction, employing a Nonius Kappa CCD diffractometer (Mo
$K\alpha$, $\lambda=0.71073$ \AA), were used to check the crystal
structure and phase purity of our samples. No indication of any second
phases in the samples was detected and the X-ray results confirmed
that the crystal structure is a base-centered monoclinic structure,
space group $C2/m$ as shown in Fig.~\ref{fig:struct}. The cell volume,
585.06(16) \AA$^3$, and lattice parameters, $a=10.0979(15)$ \AA,
$b=7.6670(15)$ \AA, and $c=7.8733(10)$ \AA\hspace{0.01in} with $\beta
= 106.298(7) ^\circ$, match previous measurements well. The single
crystal X-ray measurements showed the $c$-axis to be aligned with the
longest of the crystal dimensions and further details of these
measurements are included in the Supplementary Materials.  Crystals
were annealed between 500 and 650 $^\circ$C in an evacuated, sealed,
fused silica tube and are compared to the results of a crystal that
was sealed in a silica tube containing 1 atm of air and annealed at
550 $^\circ$C to check for changes in the magnetic properties due to
oxidation. No discernible differences were found between the crystals
annealed in air and those annealed in vacuum.

%figure 0 Crystal Structure

A Quantum Design (QD) MPMSXL SQUID magnetometer equipped with a 5-T
superconducting magnet was used to measure the magnetic
susceptibility, $\chi$, and magnetization, $M$, of the crystals from 2
to 800 K.  The $M$ and $\chi$ reported here have not been corrected
for the effects of demagnetizing fields because of the difficulty
associated with accurately determining demagnetization factors for
oddly shaped single crystals. Comparisons with previous measurements
performed on polycrystalline samples indicate that such corrections
are not significant\cite{kawamiya1,kawamiya2}. Specific heat
measurements were performed in a QD PPMS using a standard heat pulse
technique from 2 to 400 K in zero field and with a slope-analysis
method in fields up to 0.5 T in the vicinity of $T_1$. The specific
heat data were carefully corrected by subtracting the contribution
from the measurement addenda.  The electrical resistivity and Hall
effect measurements were performed on rectangular-shaped samples
polished with emery paper. Thin Pt wires were attached to four Epotek
silver epoxy contacts with an average spacing between the voltage
probes of 0.3 mm. The resistivity, magnetoresistance, and Hall effect
measurements were performed at 19 Hz using standard lock-in techniques
in a gas flow cryostat and a 5-T superconducting magnet. Hall effect
measurements were corrected for any misalignment of the leads by
symmetrizing the data collected at positive and negative fields.

\section{Experimental Results}
\subsection{Magnetic Properties of Fe$_3$Ga$_4$}
\begin{figure}[htb]
  \includegraphics[angle=90,width=3.2in,bb=0 0 285
  415,clip]{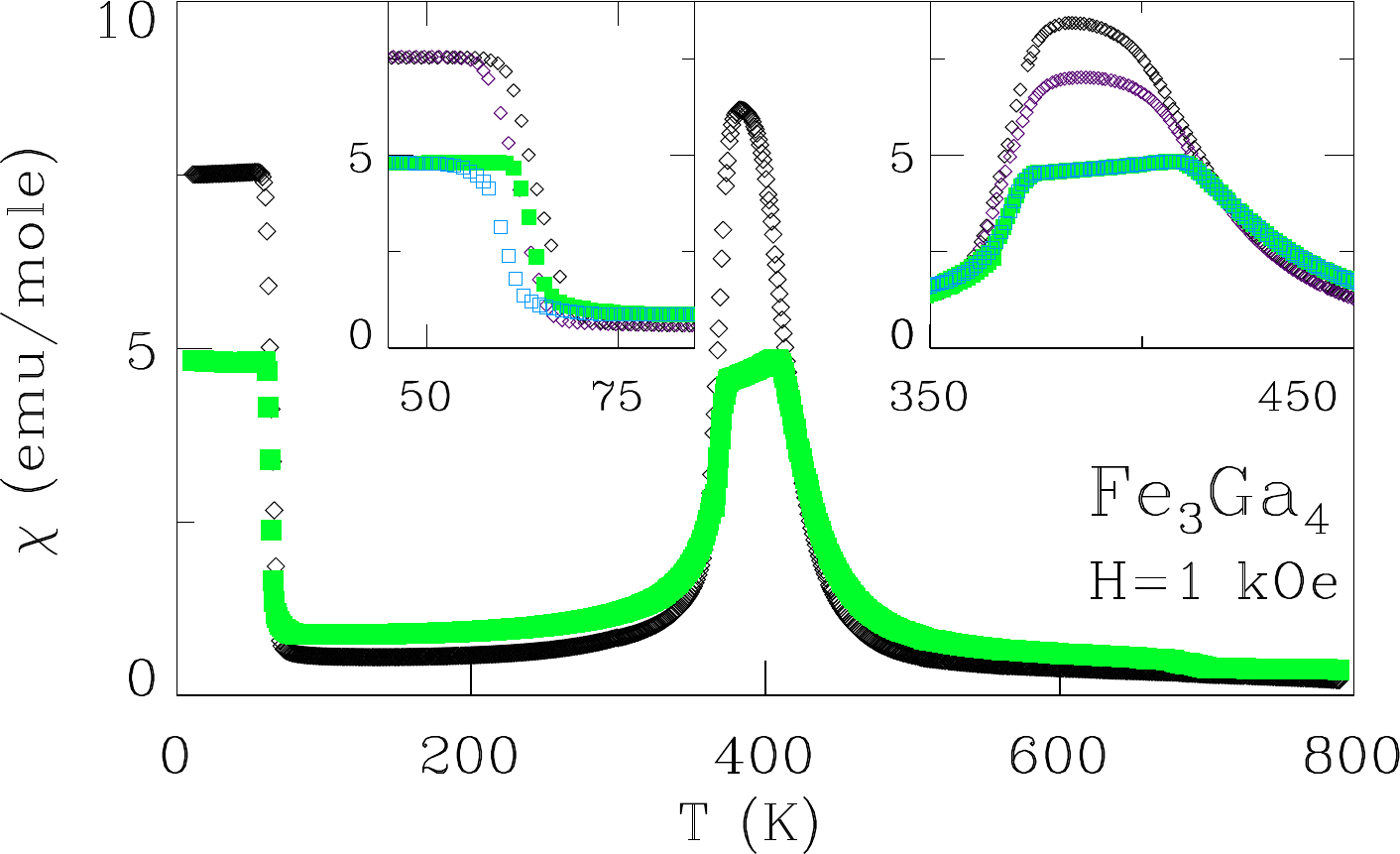}%
  \caption{\label{fig:chi} Magnetic Susceptibility. Magnetic
susceptibility of Fe$_3$Ga$_4$ in a field of 1 kOe oriented parallel
(black diamonds) and perpendicular (green squares) to the $c$-axis of
the plate shaped crystals. Insets: Hysteresis observed upon cooling
(purple diamonds and blue squares) and warming (black diamonds and
green squares) in the temperature regions of the phase transitions.
Note that in the temperature range between 350 and 450 K the warming
and cooling scans result in nearly indistinguishable $\chi$ for the
case of the magnetic field perpendicular to the $c$-axis making the
difference between the blue and green squares difficult to discern in
right-most inset. }
\end{figure}

The magnetic susceptibility of a Fe$_3$Ga$_4$ single crystal is
presented in Fig.~\ref{fig:chi} for two orientations of the magnetic
field with respect to the crystallographic $c$-axis. There are several
magnetic phases and phase transitions evident in our data including a
sharp change in $\chi$ at $T_1=68$ K. Above $T_1$ there is a wide
temperature region of smaller $\chi$ which evolves into a second
region of large $\chi$ above $T_2=360$ K.  At temperatures above
$T_3=420$ K the susceptibility is substantially reduced.  We have
indicated the hysteresis observed for increasing and decreasing
temperatures in the regions surrounding $T_1$ and $T_2$ in the
insets. A first order transition at $T_1$ is indicated by the
substantial hysteresis in this temperature range while hysteretic
behavior is not clearly indicated at the higher temperature
transitions. The phase transitions identified here match well those
that were previously identified in polycrystalline
samples\cite{wagini1,kawamiya1,kawamiya2,duijn,duijnthesis}. 

In addition to these more obvious features, we also observe a small
decrease in $\chi$ above $T_4=685$ K that was noticed in very early
investigations of the properties of Fe$_3$Ga$_4$\cite{wagini1} but
ignored in subsequent treatments. This earlier work concluded that
Fe$_3$Ga$_4$ was magnetic below 697 K with a transition near 420 K to
a second, higher moment, ferromagnetic phase\cite{wagini1}. At this
point it is not apparent if this feature in $\chi(T)$ is indicative of
a subtle magnetic transition, a structural or electronic transition,
or results from a small amount of a second phase. However, we have no
evidence for a second phase within the samples from the X-ray
diffraction investigations.  We note that there are a few possible Fe,
Ga, and O compounds that, if present, could provide a magnetic signal
with a Curie temperature in this range.  These include metastable
Fe$_{1-x}$Ga$_x$, a dilution of bcc Fe with Ga, which has a Curie
point near 685 K for $x\sim 0.26$\cite{wagini2}. In addition,
Fe$_{3-x}$Ga$_x$O$_4$ with $x=0.5$ is a possible impurity phase since
its Curie temperature is also close to $T_4$\cite{gamari,moglestue}.
We point out, however, that multiple crystals grown under different
conditions and annealing histories all displayed a similar signal,
including those annealed in either vacuum or in air, and that the
magnetic signal displays a substantial anisotropy below $T_4$. Thus,
we consider it somewhat unlikely that Fe$_{1-x}$Ga$_x$ or
Fe$_{3-x}$Ga$_x$O$_4$ impurities in the crystals would all contain the
same Ga dilution level resulting in a magnetic transition at $T_4$ in
all samples measured over this temperature range. Attempts to fit a
modified Curie-Weiss form to the data above 500 K were not
satisfactory as the data are poorly represented by a simple
paramagnetic response. This conclusion is consistent with earlier
measurements on polycrystalline samples to higher temperatures where a
Wiess temperature of 720 K and a fluctuating moment of $J=0.75$ (where
the authors have assumed a $g$-factor of 2) were
determined\cite{kawamiya2}. Further investigation of the structural
and electronic properties of Fe$_3$Ga$_4$ in proximity to $T_4$ are
necessary to resolve the cause of the change in the magnetic
susceptibility we observe.

%figure 2 magnetic susceptibility

\begin{figure}[htb]
  \includegraphics[angle=90,width=3.2in,bb=0 0 475
  540,clip]{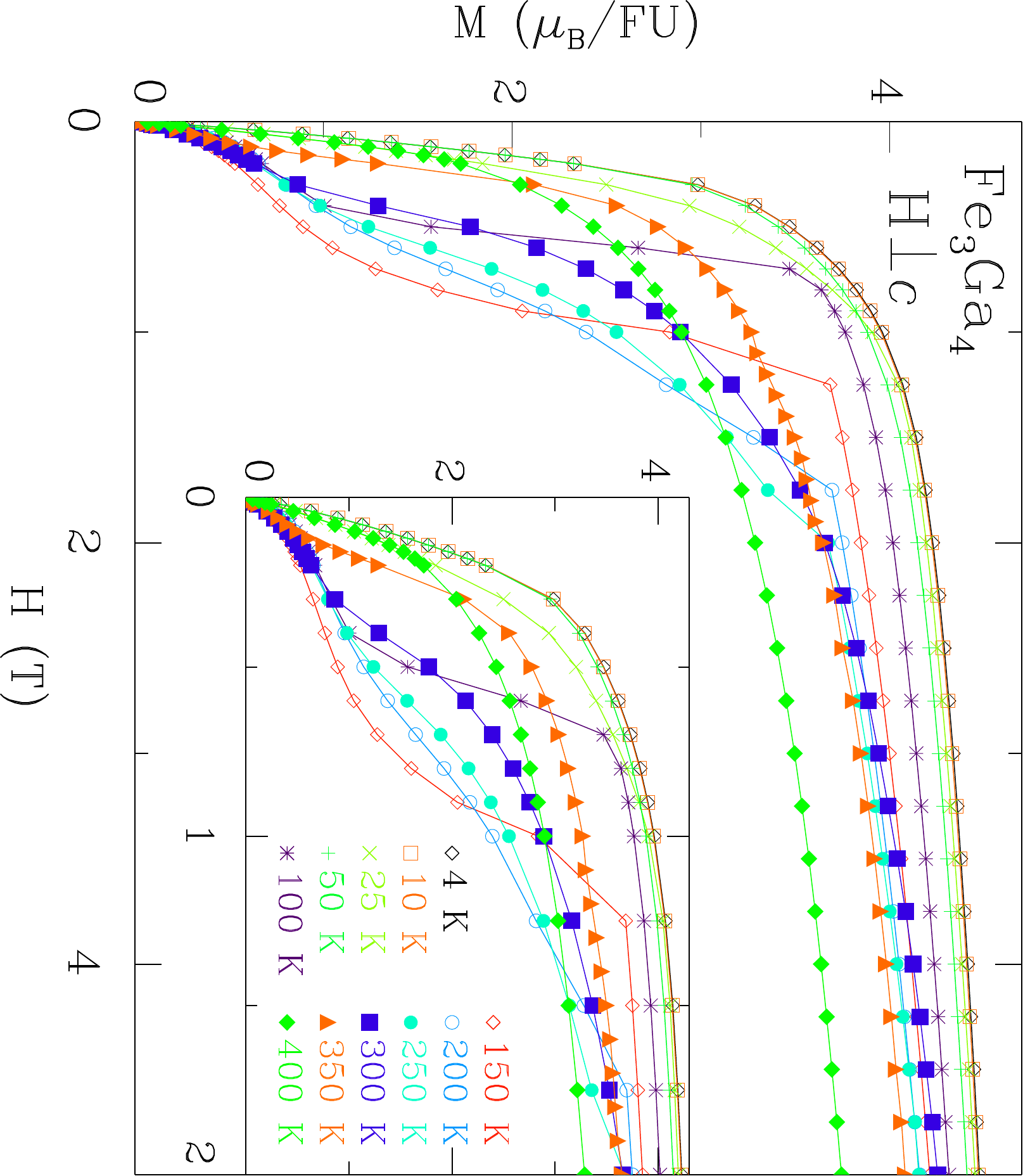}%
  \caption{\label{fig:m1} Magnetization. Magnetization, $M$, of
single-crystal Fe$_3$Ga$_4$ measured with a field, $H$, oriented
perpendicular to the $c$-axis of the crystal.  Temperatures are
indicated in the figure. Inset: Low field $M(H)$ plotted to highlight
the rapid changes to $M$ below 2 T. }
\end{figure}

\begin{figure}[htb]
  \includegraphics[angle=90,width=3.2in,bb=0 0 475
  540,clip]{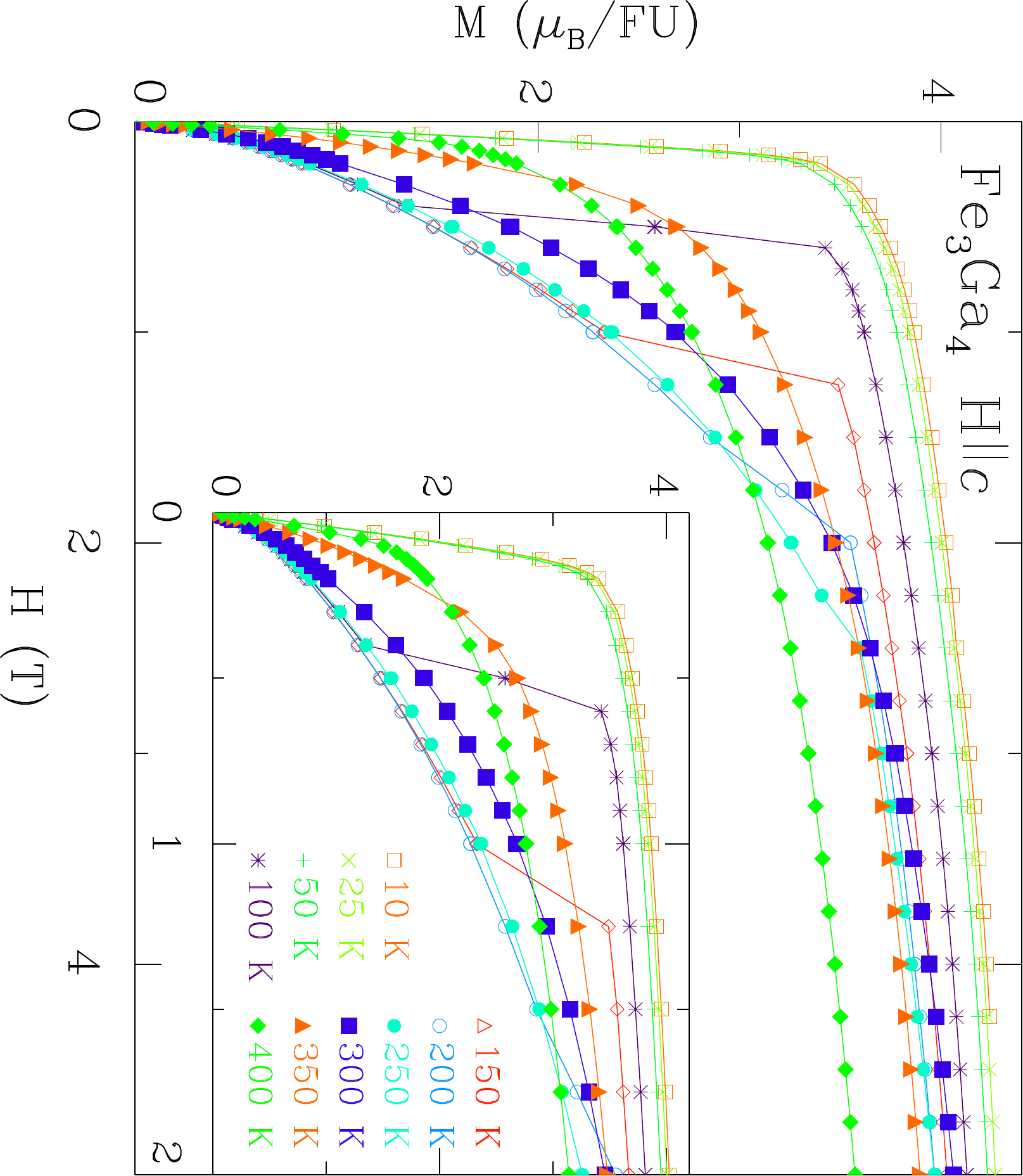}%
  \caption{\label{fig:m2} Magnetization, $M$, parallel to the
$c$-axis. Same sample as in {\protect\ref{fig:m1}}. Temperatures are
indicated in the figure. Inset: Low field $M(H)$ plotted to highlight
the rapid changes to $M$ below 2 T.}
\end{figure}

To better understand the nature of the different magnetic states
identified in Fig.~\ref{fig:chi}, we measured the isothermal
magnetization at several temperatures between 4 and 400 K as displayed
in Figs.~\ref{fig:m1} and \ref{fig:m2}. These data demonstrate both
the similarities in $M(H)$ for $T<T_1$ and $T> T_2$ K as well as the
metamagnetic transition for intermediate $T$'s. The metamagnetic
transition is observed to be particularly sharp for $H$ parallel to
the $c$-axis. Having single crystals has also allowed us to
characterize the anisotropy in $H_{mm}$.  As noticed in earlier
investigations, the critical field for this transition increases with
$T$. We observe this unusual trend for both field orientations below
150 K. While $H_{mm}$ continues to increase for fields parallel to the
$c$-axis, the magnetization step size decreases with $T$. For fields
perpendicular to the $c$-axis $H_{mm}$ decreases for $T$ above 150
K. We note that in this orientation, the data display two transitions
indicating that the crystal was somewhat misaligned with the correct
crystallographic orientation to observe only a single transition.  The
saturation magnetization, $M_S$, at low $T$ corresponds to 1.5 $\mu_B$
per Fe somewhat larger than the magnetic moment estimated from
M\"{o}ssbauer measurements (1.38 $\mu_B$/Fe)\cite{kawamiya1} and that
seen in the previous $M_S$ measurements of polycrystalline samples
(1.17 to 1.27 $\mu_B$/Fe)\cite{kawamiya2,duijnthesis}.

%figure 3 magnetization 1
%figure 4 magnetization 2

\begin{figure}[bht]
  \includegraphics[angle=90,width=3.2in,bb=0 0 285
  395,clip]{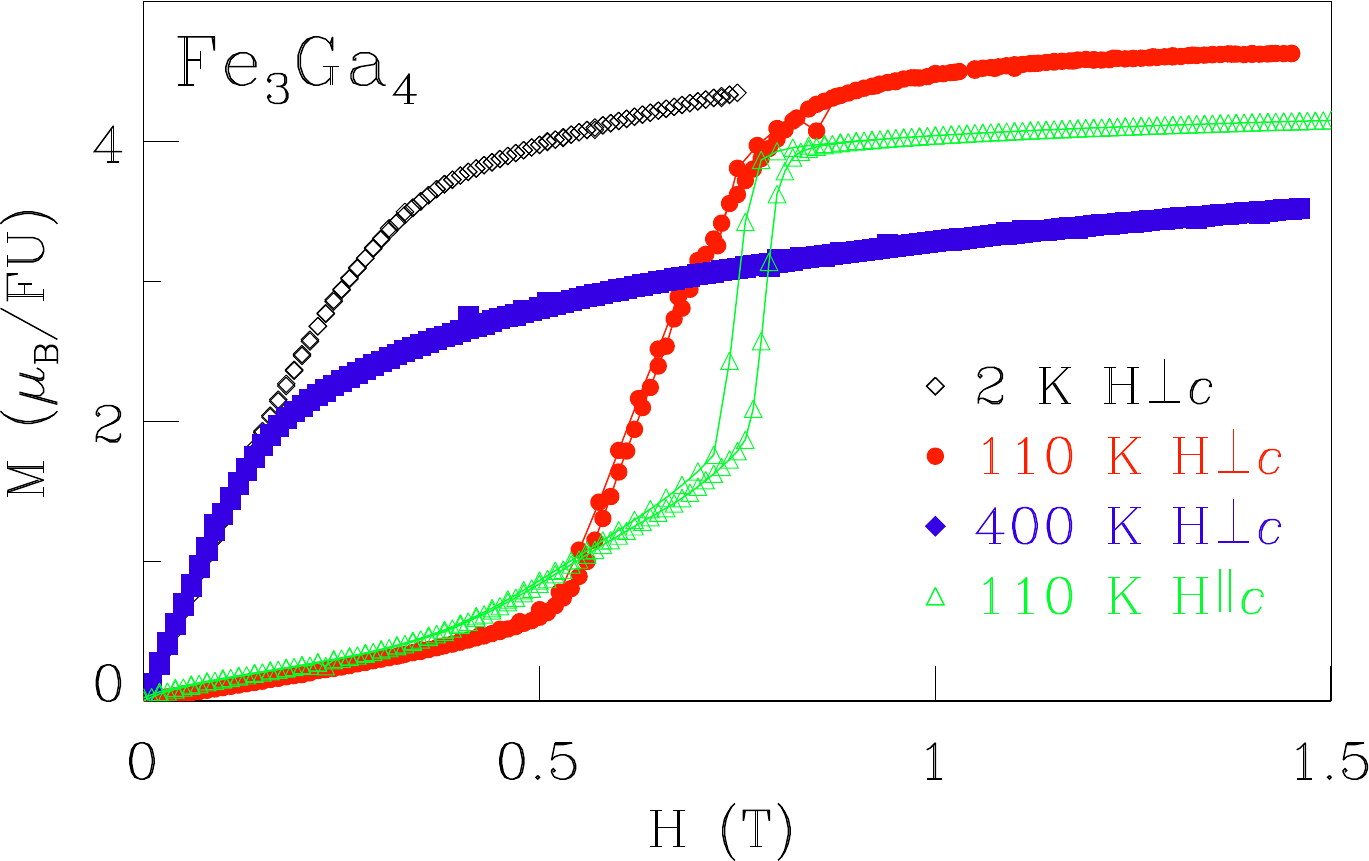}%
  \caption{\label{fig:m3} Low Field Magnetization.  Magnetization
displayed in two different field orientations with respect the $c$-axis. 
Data shown at three temperatures to demonstrate the
variations that occur in the three distinct magnetically ordered
regions that we have identified. }
\end{figure}

The magnetization, its anisotropy, and the hysteresis associated with
the metamagnetic transition in a crystal grown at a later time via
iodine vapor transport and treated in the same manner are demonstrated
in Fig.~\ref{fig:m3}. A hysteresis of about 400 Oe is evident when the
crystal is oriented with its $c$-axis parallel to the magnetic field
while a much smaller history dependence is observed when the crystal
is rotated so that $H$ is perpendicular to the $c$-axis. The
similarity of the magnetic response at 2 K and 400 K is also displayed
in the figure which motivates our identification of a FM phase below
$T_1$ and between $T_2$ and $T_3$.  However, we note that
demagnetization effects may reduce the differences apparent in these
curves and that the temperature dependence of $\chi$ is not that of a
prototypical PM-to-FM transition above 350 K. Thus, it remains a
possibility that the magnetic state between $T_2$ and $T_3$ is more
complex, such as a canted AFM state with a FM component making neutron
diffraction experiments essential to resolving the magnetic state in
this $T$ range.

%%figure 5 magnetization plot 3 low fields

The variation of the magnetic transition temperatures of the samples
with annealing mentioned above suggests that the magnetic state is
sensitive to subtle changes to the stoichiometry, disorder, or crystal
structure of the samples. We observed that the as-grown samples can
have substantially reduced values of $T_1$, $T_2$, and $T_3$, but
annealing at 550 $^\circ$C was sufficient to increase these transition
temperatures so that they more closely matched those found in the
previously measured polycrystalline
samples\cite{wagini1,kawamiya1,duijnthesis}.

\subsection{Specific Heat of Fe$_3$Ga$_4$}

\begin{figure}[bht]
  \includegraphics[angle=90,width=3.2in,bb=0 0 475
  580,clip]{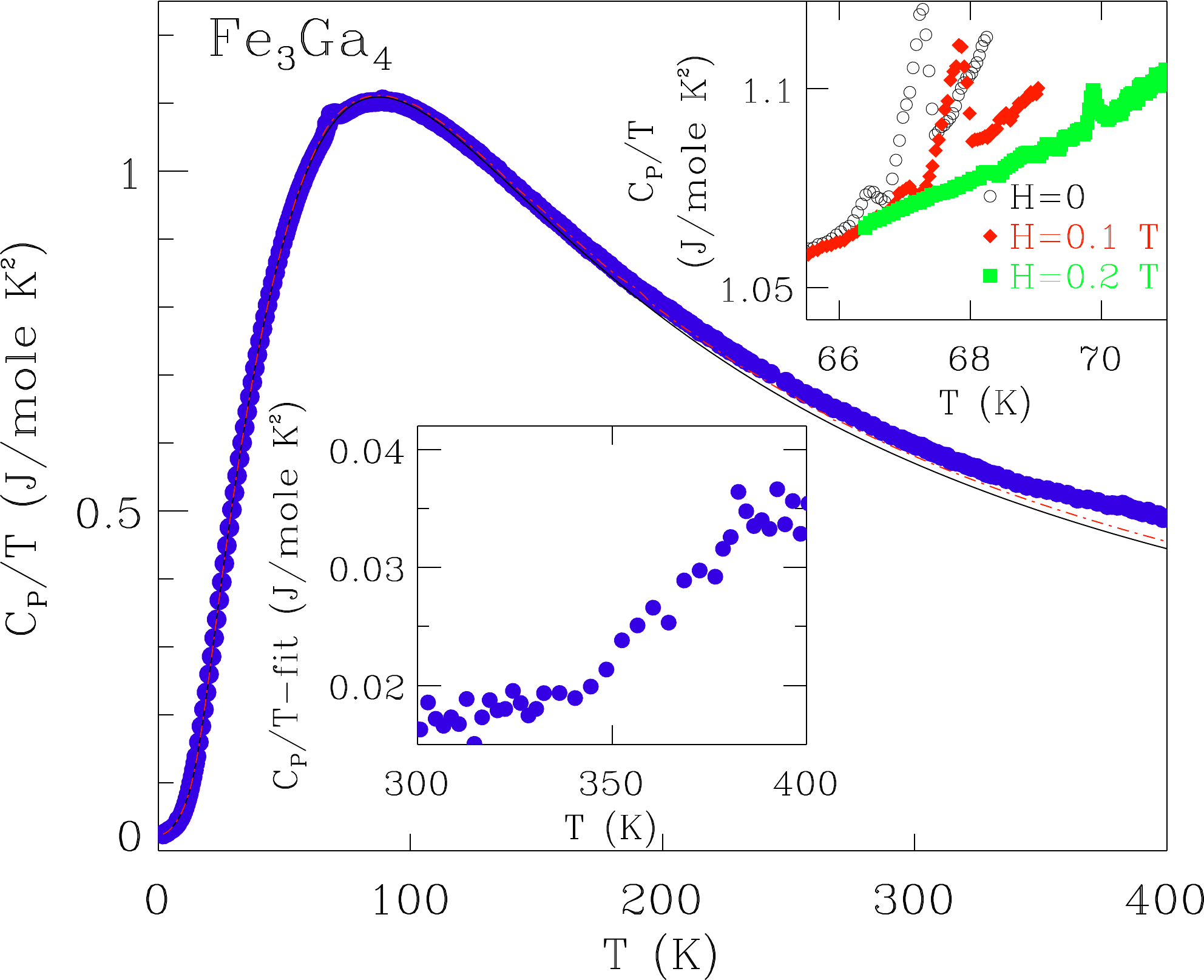}%
  \caption{\label{fig:spheat} Specific Heat. Specific heat, $C_P$
divided by temperature, $T$, as function of $T$. Solid line represents
a fit of a model of the phonon contribution to the specific heat, see
text for details. Dashed-dotted line includes a correction to the
model to better represent the heat capacity at constant pressure based
upon the thermal expansion and
compressibility{\protect{\cite{duijnthesis}}}. Upper inset: $C_P$ as
measured by a slope analysis method between 66 and 71 K at magnetic
fields, $H$ identified in the figure. Lower inset: the difference
between the measured $C_P$ (at $H=0$) and the model for the phonon
contribution, fit, between 300 and 400 K demonstrating a substantial
increase for $T>350$ K. }
\end{figure}

The specific heat, $C_P$, of a single crystal of Fe$_3$Ga$_4$ is shown
in Fig.~\ref{fig:spheat} where $C_P/T$ is displayed between 2 and 400
K. There are two features of these data that are likely of magnetic
origin. The first is a small peak near $T_1$ and the second is the
shoulder seen above 350 K. To better understand the likely
contributions to $C_P(T)$ we have fit a simplified model of the
phononic and electronic contributions to $C_P(T)/T$ to the data that
is represented in the figure by the solid line. This model contains a
Debye and three separate Einstein modes to represent the complicated
phonon density of states of Fe$_3$Ga$_4$. It also includes a linear
temperature dependent term to model the electronic contributions which
can dominate $C_P$ at low $T$, although magnons terms may also
contribute. We found that including a larger number of independent
Einstein terms did not significantly increase the quality of the fit,
so we report this minimal model to describe the data. The best fit
value of the parameters included a Debye temperature $\Theta_D = 120$
K, Einstein temperatures of $\Theta_E= 135$, 260, and 365 K, and a
linear-in-temperature coefficient, $\gamma$, of 21 mJ/mole K$^2$. We
have also included in our modeling an estimated correction
(dashed-dotted line) to account for the difference between $C_P$ and
the heat capacity taken at constant volume. This correction is based
upon the thermal expansion and compressibility of Fe$_3$Ga$_4$ as
reported in Ref.~\cite{duijnthesis}. The estimated additional
contribution due to the thermal expansion of the sample can be seen
above $\sim150$ K.

\begin{figure}[bht]
  \includegraphics[angle=90,width=3.2in,bb=0 0 475
  580,clip]{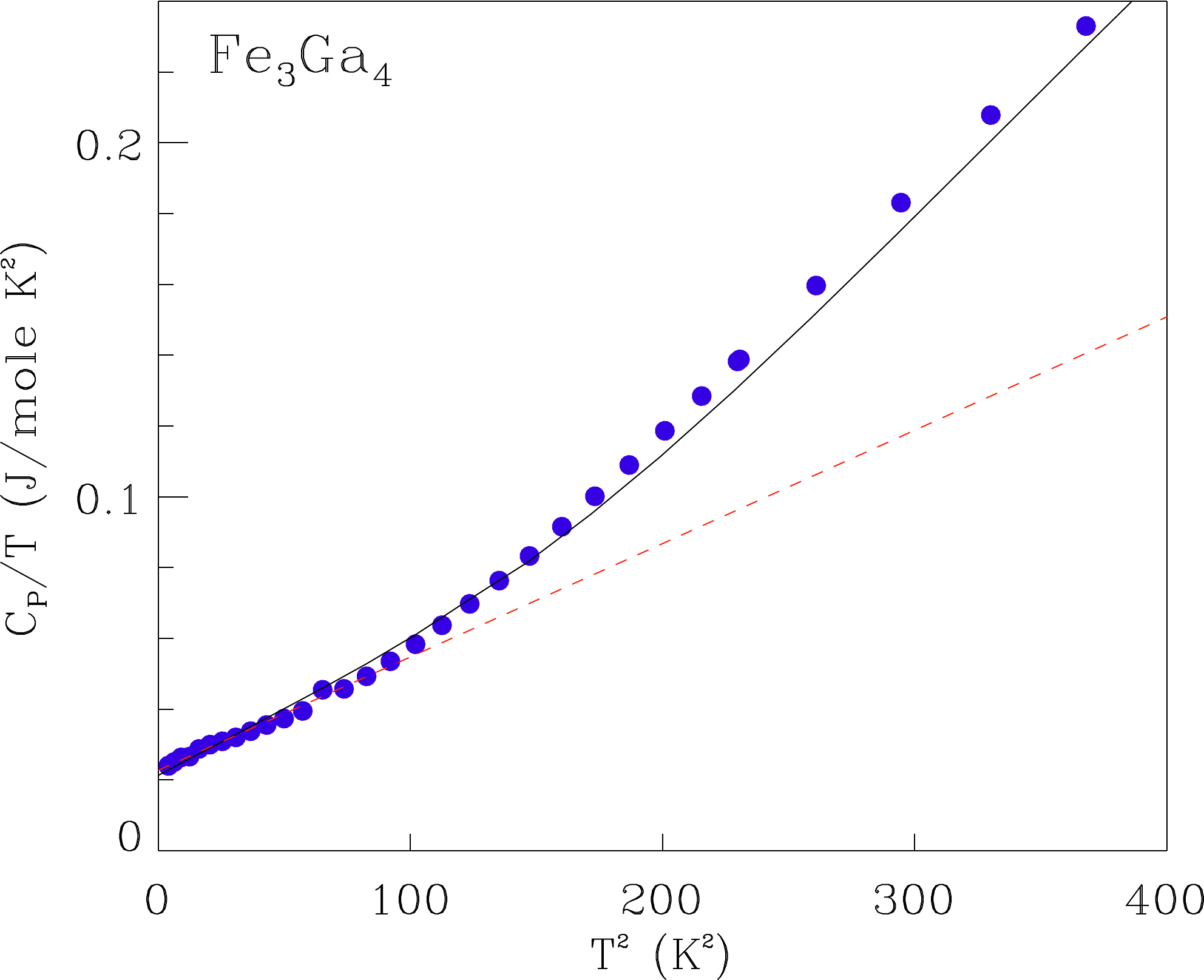}%
  \caption{\label{fig:ltspheat} Low Temperature Specific
Heat. Specific heat, $C_P$ divided by temperature, $T$ as function
$T^2$ below 20 K. Solid line represents the same fit of a model of the
phonon contribution to the specific heat as in
Fig.~{\protect{\ref{fig:spheat}}} while the dashed line represents a
fit of the simple model $C_P(T)/T = \gamma T + \beta * T^2$ to the
data below 10 K. }
\end{figure}

In Fig.~\ref{fig:ltspheat} we display the same $C_P(T)$ data at $T<20$
K using the standard form for exploring the low $T$ specific heat of
solids by plotting $C_P(T)/T$ as a function of $T^2$. Here we plot
the results of the fitting procedure described above represented by
the solid line in the figure. We have also included a fit of a linear
dependence between 2 and 10 K to represent the standard $C_P/T =
\gamma + \beta T^2$ form, dashed line. The best fit value for $\gamma$
is 23 mJ/mole K$^2$ and the $\beta$ value found corresponds to
$\Theta_D=125$ K in reasonable agreement with the more complex model
described above giving us confidence that our fitting procedure gives
us a good impression of the lattice and conduction electron
contributions to the specific heat.

%%figure 6 Specific Heat

%%figure 7 low temperature Specific Heat

The two features of the data that we suggested above are of magnetic
origin are clearly not described by our models shown in
Fig.~\ref{fig:spheat}. The first, the sharp peak at $T_1$, is highly
sensitive to $H$ as we demonstrate in the upper inset to
Fig.~\ref{fig:spheat}. Here the application of magnetic field is seen
to drive the transition to higher $T$ and to decrease the size of the
anomaly such that by 0.5 T we find no indication of a sharp anomaly in
$C_P(T)$ in this $T$ range. These data were taken with a
slope-analysis method which makes use of the measured change in the
$T$ of the sample platform over small intervals of $T$ during a
warming or cooling cycle to give greater sensitivity near sharp phase
transitions. Our model of the phonon contributions also fails to
properly fit the $C_P(T)$ data above 150 K with the difference between
data and model growing slowly until 350 K where a steep increase in
this difference is apparent in the lower inset to
Fig.~\ref{fig:spheat}. The onset of this contribution near 350 K
corresponds well with $T_2$ identified in $\chi(T)$ and, thus, we
identify this anomaly with this transition.  The entropy, $S$,
associated with the transition at $T_1$ found using the relation
$\Delta S = \int C_P(T)/T dT$ is small, $\sim 17$ mJ/mole K, or $\sim
0.2$ \% of $R \ln(2J+1)$, where we have made use of the estimated
average value for $J=0.75$ from $M_S$ (Figs.~\ref{fig:m1} and
\ref{fig:m2}). Thus, the transition from the low $T$ FM state to the
tentatively identified AFM state at $T_1$ does not involve a large
entropy change. We have also estimated $\Delta S$ associated with the
rise in $C_P(T)$ above $T_2$ finding $\Delta S = 0.43$ J/mole K
between $T_2$ and 400 K providing an upper bound to the magnetic
entropy change.

\begin{figure}[bht]
  \includegraphics[angle=90,width=3.2in,bb=0 0 475
  580,clip]{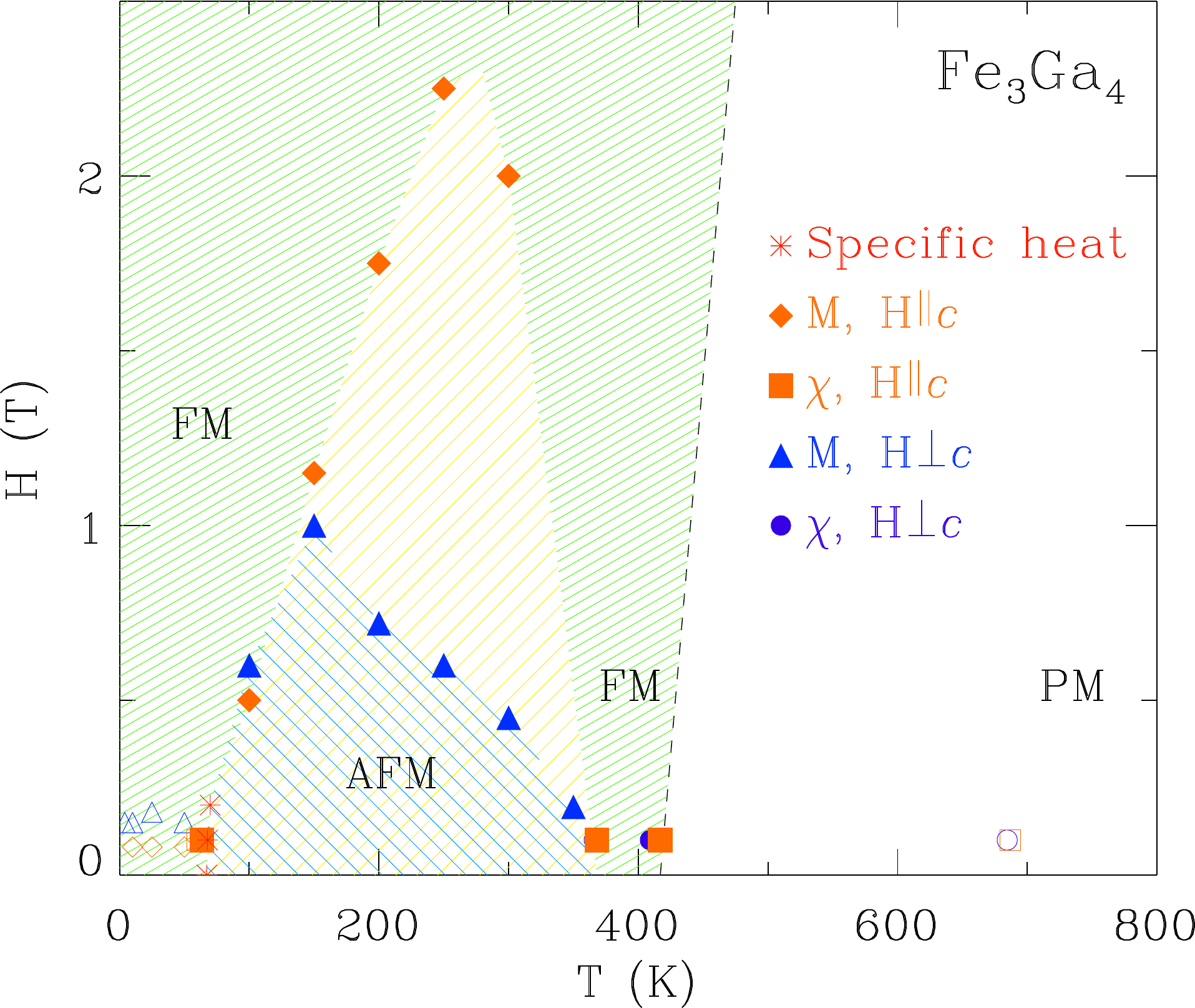}%
  \caption{\label{fig:phasd} Phase Diagram. Temperature, $T$,and
magnetic field, $H$, phase diagram of Fe$_3$Ga$_4$ based on
magnetization, $M$, susceptibility, $\chi$, and specific heat
measurements. Symbols represent the phase boundaries determined by the
maxima of the $H$-derivatives of the isothermal $M$, maxima of the $T$
derivatives of $\chi$, and maxima in $T$ dependence of the specific
heat. These phase boundaries are plotted for $H$ in both the $H\perp
c$ and the $H\parallel c$ configurations as indicated in the
figure. Zero field magnetic phases are tentatively assigned as
ferromagnetic, FM, antiferromagnetic, AFM, or PM, as indicated in the
figure. Data at 685 K indicate $T_4$ where a weak peak in $d\chi/dT$
is observed which may indicate a phase transition, but that has not
been fully characterized. }
\end{figure}

A tentative phase diagram based upon our $M(H,T)$ and $C_P(T)$ data is
presented in Fig.~\ref{fig:phasd} to demonstrate the overall behavior
that we have observed. A phase diagram based on polycrystalline
measurements can be found in Ref.~\cite{duijnthesis}. Here, we have
employed the earlier designations for the phases that were assigned as
ferromagnetic and antiferromagnetic, but these are also only tentative
as neutron scattering experiments to date have been
inconclusive\cite{duijnthesis}. The anisotropy in $H_{mm}$ above 150 K
is evident and defines a large portion of the phase diagram.  The open
symbols designate fields where we observe changes in $dM/dH$ which may
indicate a rearrangement of magnetic domains in the FM phase.  In
Fig.~\ref{fig:phasd} we have also indicated $T_4$ where we observe a
small increase in $\chi$. The dashed line in the figure is merely a
designation of the crossover between the FM and PM state at finite
field that is poorly defined and that we have not adequately explored.

%%figure 8 phase diagram

\subsection{Electrical Transport Properties of Fe$_3$Ga$_4$}

\begin{figure}[bht]
  \includegraphics[angle=90,width=3.2in,bb=0 0 275
  425,clip]{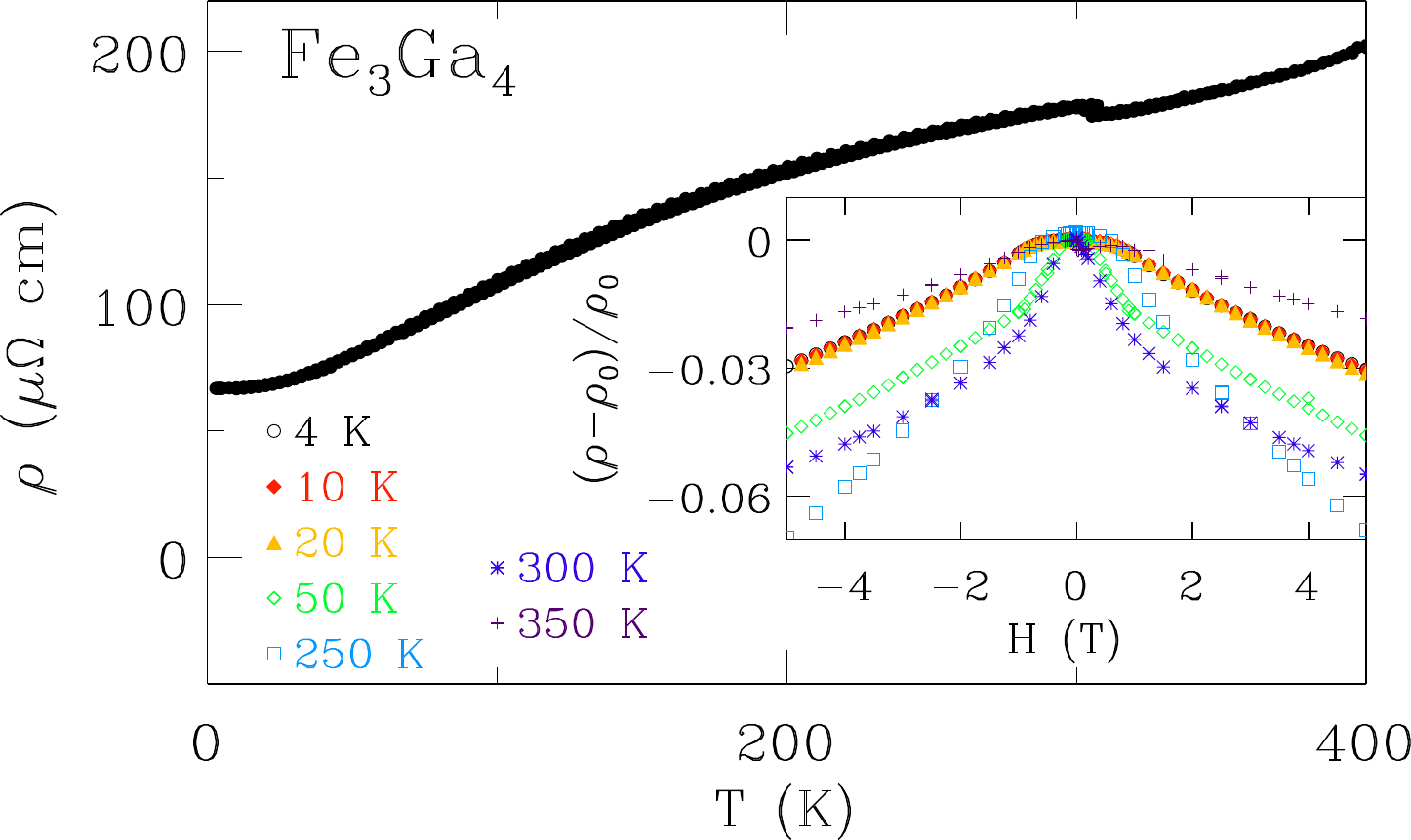}%
\caption{\label{fig:res} Resistivity and Magnetoresistance. The
zero-field resistivity, $\rho$, as a function of temperature,
$T$. Inset: Magnetoresistance, $(\rho(H) - \rho_0)/\rho_0$, where $H$
is the magnetic field and $\rho_0$ is the zero-field resistivity, at
temperatures indicated by the key in the main frame.  }
\end{figure}

The electrical resistivity, $\rho$, magnetoresistance, MR, and Hall
effect of our crystals were measured in a configuration where the
current was along the $c$-axis of the crystals and the field was
perpendicular to the direction of the current (transverse
MR). $\rho(T)$ shown in Fig.~\ref{fig:res} can be compared to previous
measurements performed on polycrystalline
samples\cite{wagini1,duijn}. The behavior is metallic with a residual
resistivity ratio (RRR) of 2.7 and a strong anomaly near $T_2$ (close
to 310 K for the crystal whose $\rho$ is shown in the
figure). Although the room-$T$ value for $\rho$ ($\sim 200$
$\mu\Omega$ cm) is smaller than in previous measurements, the RRR is
significantly smaller than that found by Duijn\cite{duijnthesis}.  No
easily identifiable anomaly near $T_1$ was observed. In agreement with
Ref.~\cite{duijnthesis} we find a low $T$ $\rho$ that is well
described by a $T^2$ dependence consistent with the moderately
enhanced $\gamma$ observed in $C_P(T)$.

%%figure 9 resistivity and MR

We find a negative MR, $\Delta \rho / \rho_0$ where $\Delta \rho =
\rho - \rho_0$ and $\rho_0$ is the zero-field value of $\rho$, at all
temperatures investigated (between 4 and 350 K) as displayed in the
inset to Fig.~\ref{fig:res}. This is in contrast to the previously
published work on polycrystalline materials\cite{duijnthesis} where a
positive MR was found below 15 K and where a positive contribution was
apparent above 350 K.  A negative MR is to be expected for itinerant
ferromagnets where the field dependence of the carrier fluctuation
scattering can dominate, particularly near phase
transitions\cite{moriya}. The low-field negative MR appears to be at a
maximum near $T_2$ where the anomaly in the $\rho(T)$ is observed. In
addition, we also find discontinuous changes to $d\rho / dH$ at fields
close to saturation. The $\Delta \rho / \rho_0$ values at 5 T are
nearly a factor of 2 smaller than that reported by
Duijn\cite{duijnthesis} most likely due to the larger residual
resistivity of our crystals.

\begin{figure}[bht]
  \includegraphics[angle=90,width=3.2in,bb=0 0 475
  580,clip]{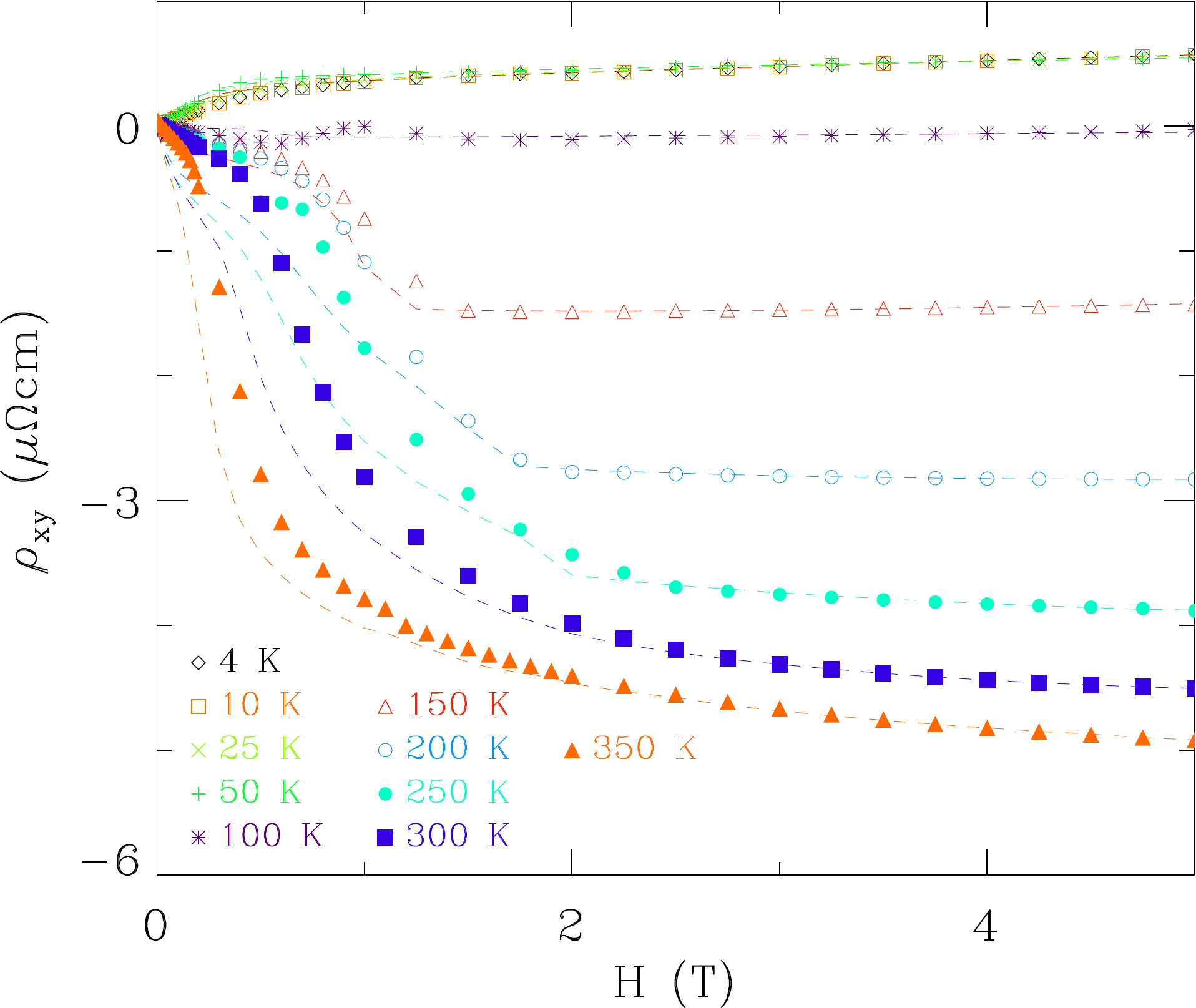}%
  \caption{\label{fig:hall1} Hall effect. The Hall resistivity,
$\rho_{xy}$ of Fe$_3$Ga$_4$ vs.\ magnetic field, $H$, at temperatures
indicated in the figure. Dashed lines are fits of a standard model of
the anomalous Hall effect, see text for details.}
\end{figure}

In magnetic materials the Hall effect is usually dominated by the
anomalous contributions, referred to as the anomalous Hall effect
(AHE), stemming from spin-orbit coupling (intrinsic) or spin orbit
scattering (extrinsic) contributions\cite{nagaosa,manyala2}.  This
expectation is met in Fe$_3$Ga$_4$ as we demonstrate in
Fig.~\ref{fig:hall1} where a Hall resistivity, $\rho_{xy}$, as large
as $-5$ $\mu\Omega$ cm at 5 T is observed. The field dependence of
$\rho_{xy}$ resembles that of $M(H)$ from the same crystal in the same
orientation presented in Fig.~\ref{fig:m1}. Besides the large
magnitude of $\rho_{xy}$ at high temperatures, the most apparent
feature is its change of sign at $\sim 100$ K.

%%figure 10 Hall effect

\begin{figure}[bht]
  \includegraphics[angle=90,width=3.2in,bb=0 0 475
  580,clip]{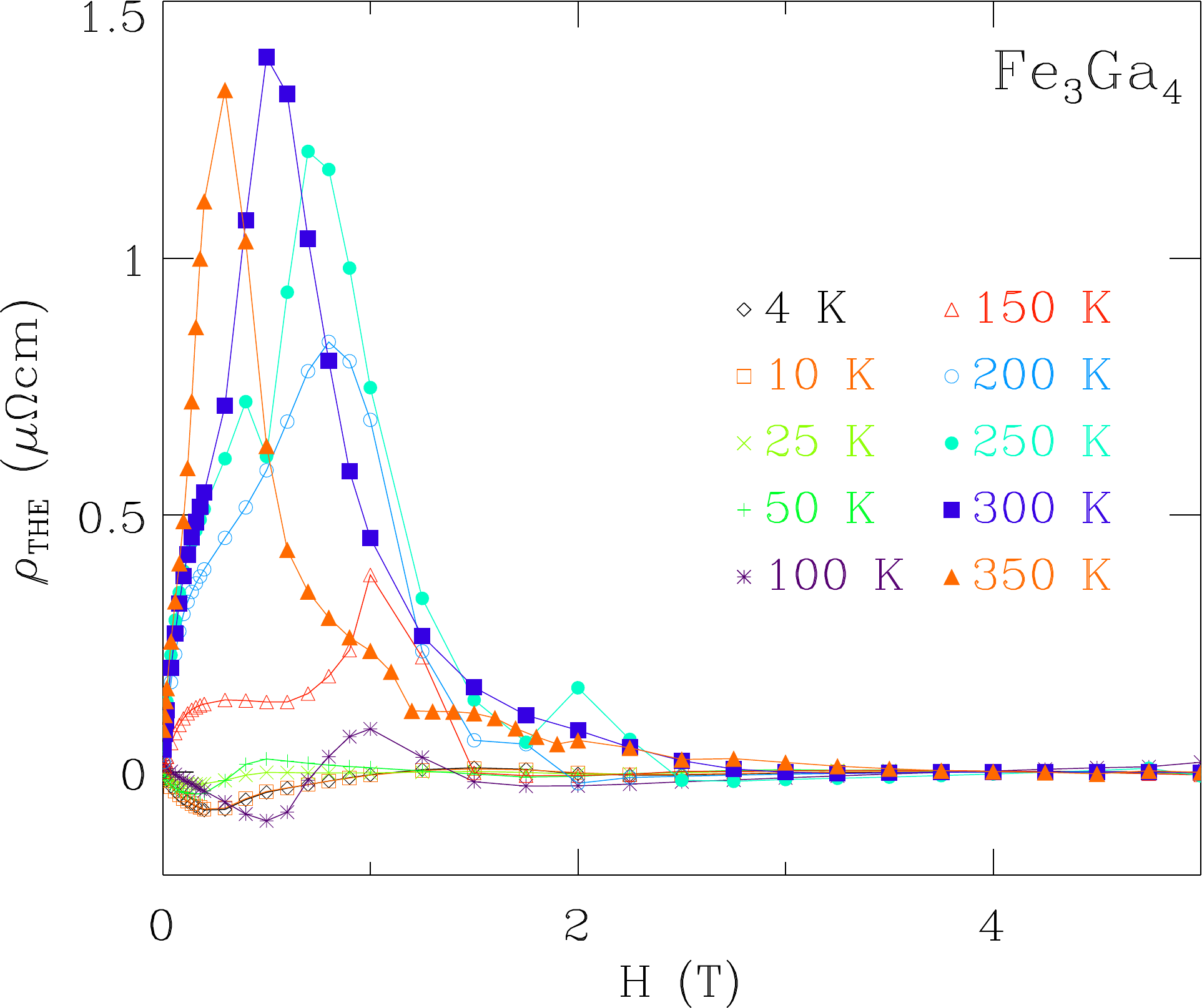}%
  \caption{\label{fig:hall2} Topological Hall Effect. The topological
contribution to the Hall resistivity of Fe$_3$Ga$_4$
$\rho_{THE}=\rho_{xy} -\rho_{xy-fit}$ - where $\rho_{xy-fit}$
corresponds to the dashed lines in Fig.~{\protect{\ref{fig:hall1}}},
vs.\ magnetic field, $H$, at temperatures indicated in the figure. }
\end{figure}

 In order to quantify the Hall constants and better understand the
relationships between the ordinary and anomalous contributions to
$\rho_{xy}$, we fit the usual model for the AHE, which is described by
the expression $\rho_{xy} = R_0H + 4\pi M R_S$, to the data. In this
model $R_0$ is the ordinary Hall constant which is closely related to
the sign and density of the conducting carriers and $R_S$ is the
anomalous Hall constant. $R_S$ has been shown to be proportional to
$\rho^2$ when the intrinsic or side-jump scattering mechanisms
dominate, and to $\rho$ when the skew-scattering mechanism is largest
(at very low resistivities). Since the intrinsic mechanism is thought
to describe materials in the resistivity range of our Fe$_3$Ga$_4$
crystals (Fig.~\ref{fig:res}), we have assumed a $R_S = S_H * \rho^2$
dependence when interpreting the anomalous term. The dashed lines
shown in Fig.~\ref{fig:hall1} are the results of this fitting
procedure, $\rho_{xy-fit}$, where the high field data were more
heavily weighted since the magnetization is near saturation and the
linear dependence of $\rho_{xy}/H$ on $M/H$ is more apparent. While
this model describes the data qualitatively well, there are distinct
differences between the data and model at low fields.  To highlight
these differences, we plot the residual Hall effect, the difference
between the data and the model, $\rho_{THE}=\rho_{xy} -
\rho_{xy-fit}$, in Fig.~\ref{fig:hall2}. It is interesting to note
that $\rho_{THE}$ is largest in the field region where $dM/dH$ is
largest, that is in the range 0.1 to 1 T. It is clear from a
comparison of the low field $\rho_{xy}$ in Fig.~\ref{fig:hall1} and
$M(H)$ in Fig.~\ref{fig:m1} that the magnetization has a low field
contribution that is missing from the AHE. There are two possible
reasons for the failure of the model to capture this low field
behavior. The first is to speculate that the low field $M(H)$ is
dominated by an extrinsic contribution most likely a magnetic second
phase that charge carriers are not sensitive to. However, this would
require a large portion of the crystals, $\sim 10$\%, to be made up of
this second phase, which is not consistent with the X-ray diffraction
data. Instead, we assert that there is likely a non-coplanar magnetic
moment at low fields in Fe$_3$Ga$_4$ so that an AHE stemming from a
topological contribution to the Hall effect, $\rho_{THE}$, is
responsible for the difference between the data and the simple
model\cite{ye,nagaosa,taguchi,machida,kanazawa}. Since there are no
reliable data determining the character of the magnetic order in
Fe$_3$Ga$_4$ we are not able to completely resolve this issue at this
time.

%%figure 11 Residual Hall effect

\begin{figure}[bth]
  \includegraphics[angle=90,width=3.2in,bb=0 0 475
  450,clip]{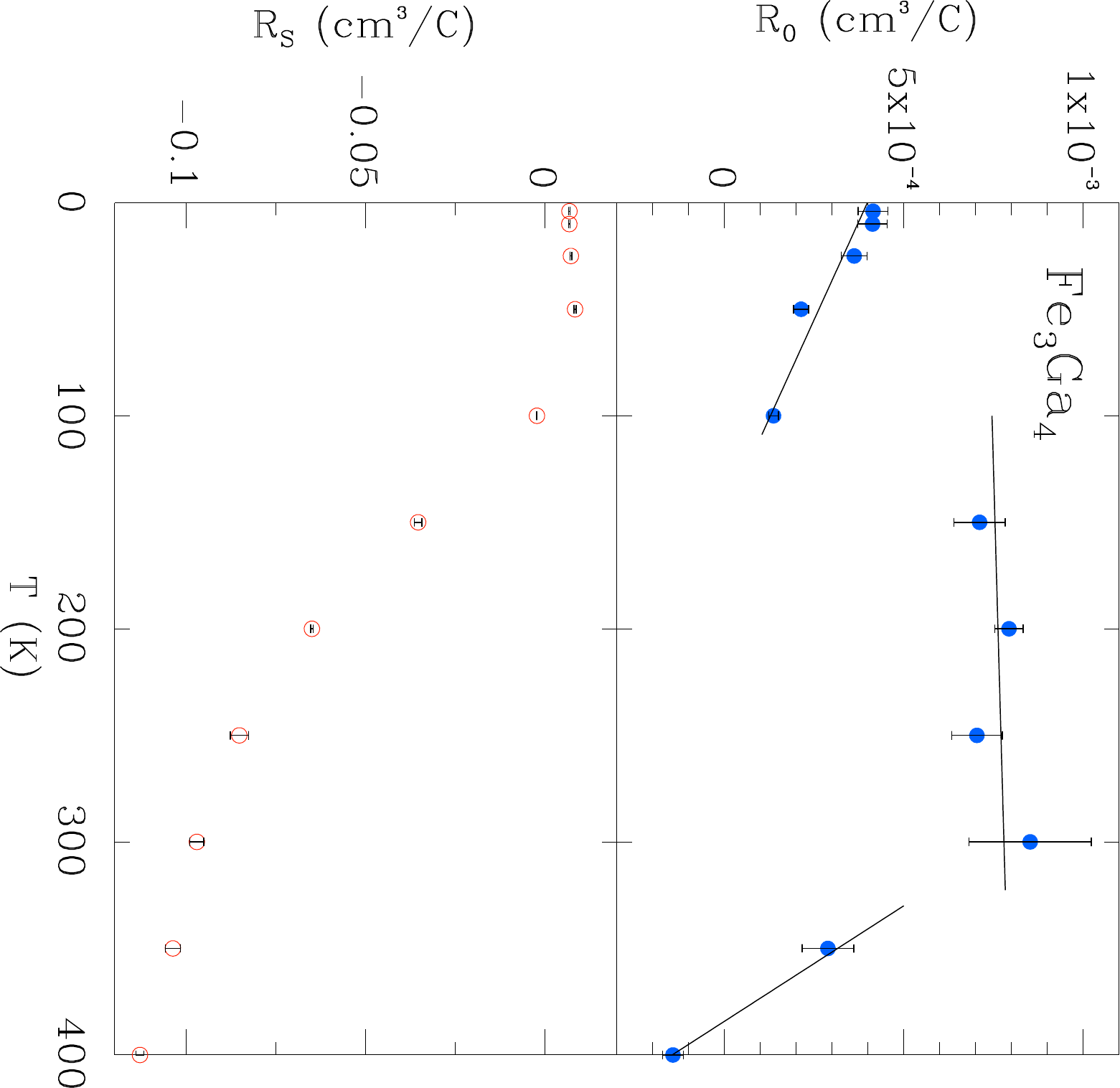}%
  \caption{\label{fig:hall3} Hall constants (a) The ordinary Hall
coefficient, $R_0$ of Fe$_3$Ga$_4$ vs.\ temperature, $T$. Lines are
guides to the eye. (b) The anomalous Hall coefficient, $R_S$,
vs. $T$. Coefficients were determined from fits of the standard model
of the anomalous Hall effect to the data as shown in
Fig.~{\protect{\ref{fig:hall1}}}.}
\end{figure}

The Hall parameters determined from our fitting procedure, $R_0$ and
$R_S$, are presented in Fig.~\ref{fig:hall3}. There are several
features of these data that are striking. While the ordinary Hall
constant is positive, hole-like carriers, below room temperature with
a value consistent with a single carrier per Fe$_3$Ga$_4$ formula
unit, $R_S$ is large ($\sim -0.1$ cm$^3$/C near room temperature and
$\sim 0.01$ cm$^3$/C at low temperature) and undergoes a sign change
near 100 K.  The change in $R_S$ from positive at low $T$ to negative
above 100 K reflects the change of sign of $\rho_{xy}$ at this $T$ and
appears to correlate well with a strong increase of $R_0$ between 100
and 150 K as well as the change in the magnetic state of the system in
this temperature and field range (see Fig.~\ref{fig:phasd}). The
decoupling of $R_0$ and $R_S$ is highlighted by this feature of the
data and further supports our assumption that the AHE results from
intrinsic, $k$-space Berry's phase related, effects. This feature may
also indicate interesting variations to the spin-orbit coupling as the
Fermi surface evolves due to the changing magnetic state. As a point
of comparison, we estimate the parameter $S_H = R_S/\rho^2$ to be
$2.4\times 10^4$ Am/V$^2$s which is smaller than found in MnSi and
MnGe, but comparable to values found in other Fe-containing itinerant
magnets FeGe and Fe$_{1-x}$Co$_x$Si\cite{silgerahe}.

%figure12 Hall constants

\section{Computational Results}
\begin{figure}[bth] 
 \includegraphics[angle=0,width=3.2in,bb=35 20 700 
 500,clip]{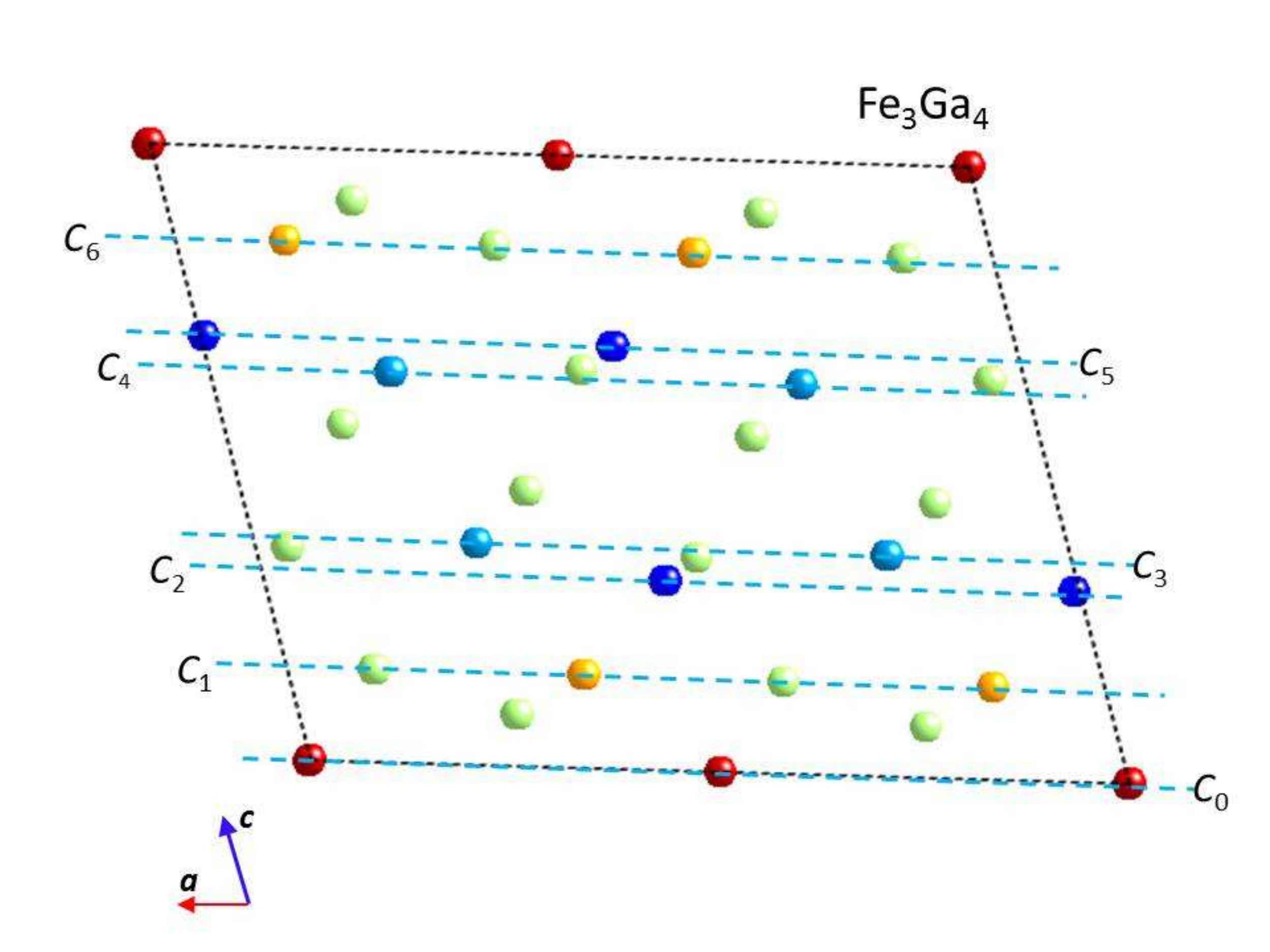}% 
 \caption{\label{fig:Estruc2} Crystal structure of Fe$_3$Ga$_4$ viewed
along the $b$-axis. Atoms are identified by their color in the same
way as in Fig.~{\protect{\ref{fig:struct}}}. Each plane of Fe atoms is
identified by the dashed lines and labeled $C_0$ through $C_6$.}
\end{figure}

The experimental results presented here, as well as those obtained
previously for Fe$_3$Ga$_4$, reveal a close competition between
magnetic states resulting in magnetic phase transitions easily
accessible via variations in temperature or magnetic field.  However,
the intermediate magnetic state has proved difficult to identify and
the critical temperatures and fields appear to be sensitive to
disorder and sample preparation conditions. In addition, a visible
jump in the electrical resistivity at $T_2$ suggests that there may be
a significant change in the electronic structure as the material
enters or leaves the AFM-like phase. Thus, to gain insight into the
likely magnetic ordering as well as the mechanism creating the close
competition between magnetic states and the resulting sensitivities,
we have performed electronic structure calculations. To gain a better
understanding of the complex low symmetry monoclinic structure of
Fe$_3$Ga$_4$, we display in Fig.~\ref{fig:Estruc2} the unit cell of
Fe$_3$Ga$_4$ in the $ac$-plane labeling the planes of Fe. In this way,
we demonstrate the crystal symmetries that exist along the
$c$-axis. Further structural details are provided in the Supplementary
Materials where the crystallographic data, including the site
symmetries and position in fractional coordinates that result from our
X-ray diffraction measurements, are presented.  The four unique Fe and
Ga atom sites in the unit cell along with their multiplicity and site
symmetries are also shown in these tables. In Fig.~\ref{fig:Estruc2} the
system is presented as consisting of Fe planes aligned along the
$c$-real space translation vector starting at $C_0$ and proceeding
through $C_6$. Several of these Fe-containing planes are related by
symmetry through a mirror plane perpendicular to the $c$-axis and
through the center of the unit cell. This makes planes $C_1$ and
$C_6$, $C_2$ and $C_5$, as well as $C_3$ and $C_4$ symmetric, leaving
only the $C_0$ plane not having a partner related to it by symmetry.
In addition, in the primitive unit cell the $C_1$ and $C_6$ planes
contain two Fe atoms while the remaining planes only contain one Fe
atom.  Supplementary Materials Table~4 lists the neighboring atoms for
both the four symmetrically unique Fe and Ga atoms in the unit cell to
demonstrate the coordination and bonding of atoms in Fe$_3$Ga$_4$.
Interestingly, the Ga atoms nearest-neighbors are almost all Fe
(except Ga4, which has a single Ga nearest-neighbor atom) while the Fe
atoms neighbors are mainly Ga with Fe atom neighbors for the four Fe
sites ranging from ~17\% to 43\% in number.

%figure 13
 
Given the complex low symmetry structure combined with the competition
between magnetic states, we carried out the electronic structure
calculations using two different density functional theory based
methods in order to validate our approach.  We employed an
all-electron full-potential linear augmented plane-wave (FLAPW)
method\cite{singh2006} based on the WEIN2K software
package\cite{Blaha2001} and a plane-wave based approach that
incorporates the projected-augmented wave (PAW) method within the
Vienna-{\it Ab-Initio} Simulation Package
(VASP)\cite{kresse1,kresse2}. For both approaches the generalized
gradient approximation based Perdew-Burke-Ernzerhof (PBE)
exchange-correlation functionals was
used\cite{PhysRevLett.77.3865}. The FLAPW simulations used the
tetrahedron integration technique on a 7$\times$8$\times$6 $k$-space
mesh for the $k$-space integration, while in the VASP simulations we
used the Methfessel-Paxton of order two integration method on a
8$\times$8$\times$8 special $k$-point mesh with a Gaussian smearing
factor of 0.2. In both methods we carefully studied the convergence of
the simulations with respect to the $k$-point mesh. In addition, for
the VASP simulations, decreasing values of the Gaussian smearing
factor were investigated.  The LAPW muffin-tin sphere radii were 2.42
and 2.17 Bohr radii for Fe and Ga atoms, respectively. Careful
convergence studies for the FLAPW plane-wave basis set were carried
out by varying the $RK_{max}$, which is the product of the smallest
atomic sphere radius $R$ times the largest $k$-vector $K_{max}$ where
$K_{max}$ determines the cut-off of the plane wave expansion used to
represent the wavefunction in the interstitial region. A value of 9.0
was found to produce accurate total energies and forces. The VASP
simulations utilized PAW potentials with the Fe atoms containing 14
valence electrons $(3p^6d^74s^1)$ and Ga atoms with 13 valence
electrons $(3d^{10}4s^2p^1)$ and a plane-wave energy cutoff equal to
293.238 eV.

The total energy in the FM and AFM states was determined in our
simulations using the Fe$_{3}$Ga$_{4}$ conventional cell for both the
WEIN2k and VASP calculations. Both of these calculations yielded a FM
state that is lower in energy by $\sim 1$ eV / unit cell, a value much
larger than the relative error (a few meV / unit cell) expected in our
calculations. In addition, both approaches yielded $\sim 0.7$\% error
in all three lattice constants as compared to the experimental values.
Given the consistency between the two methods, the remainder of the
presented simulation data will be from the VASP calculations.

Although, one can obtain an AFM solution for either the conventional
or primitive cell, the energy difference between the AFM and FM states
is large with an equivalent temperature of ~12,000 K well above the
observed magnetic transitions.  The large difference in energy between
the FM and AFM states that the simulations find can be understood by
considering the symmetry of the Fe1 crystallographic sites. Unlike a
conventional cubic cell where there are no point group operators that
map like atoms from one sublattice onto the other, in this
conventional cell there are such point group operators. Thus, to break
the symmetry and create lower energy AFM states requires careful
consideration with supercells.  Because the symmetry of the unit cell
of Fe$_3$Ga$_4$ has a inversion-type point group operation about the
Fe1 site (center bottom of the structure shown in
Fig.~\ref{fig:struct}), the construction of supercells along the
$c$-direction ought to lead to a lower energy AFM state. To this end,
we have carried out AFM and FM simulations on $1\times 1 \times 2$
supercells constructed from the Fe$_3$Ga$_4$ primitive unit cell. This
supercell configuration produces an $A$-type ordering where the Fe
spin moments alternate between Fe planes along the
$c$-direction. Simulations of this arrangement produces a total Fe
moment equal to zero (with individual Fe-moments ranging from 1.76 to
2.24 $\mu_B /$ Fe) along with Ga moments that are less than $0.1
\mu_B$ and whose sum is also zero. The simulation of the FM state
yields similar sized magnetic moments yielding an average of 1.96
$\mu_B/$ Fe and ranging from 1.80 to 2.20 $\mu_B/$ Fe. These magnetic
moments are somewhat larger than the average magnetic moment
determined from $M_S$ (1.5 $\mu_B/$ Fe).

\begin{figure}[bth]
 \includegraphics[angle=0,width=3.2in,bb=55 65 640
 440,clip]{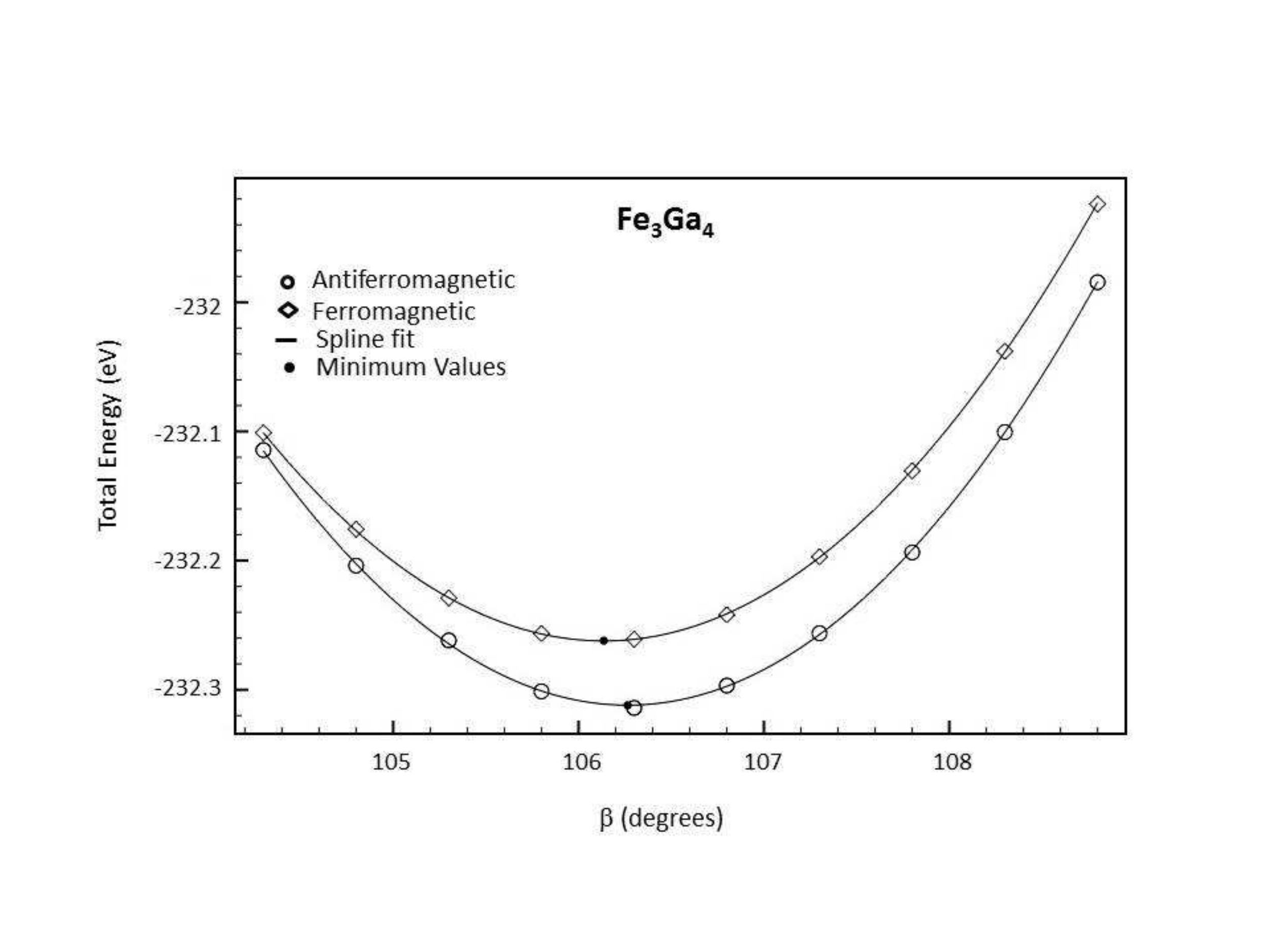}%
 \caption{\label{fig:tevsbeta} Total Energy vs.~$\beta$. The total
energy change in the calculations vs.~the structural angle $\beta$
comparing the ferromagnetic (FM) and antiferromagnetic (AFM) states.
We find that the AFM state is lower in energy at all $\beta$ near the
energy minimum and that the energy minima occur at very similar
$\beta$.}
\end{figure}

Fig.~\ref{fig:tevsbeta} displays the total energy vs.\ $\beta$ for the
$1\times 1 \times 2$ supercell in both FM and AFM states. For this
 AFM configuration we find that $\Delta E_{FM-AFM} \sim -0.031$
eV/super cell at a $\beta=106.3^{o}$.  Although this energy difference
has an equivalent temperature of ~360 K, which is in good agreement
with one of the experimentally observed AFM/FM transition
temperatures, the simulations produce an AFM ground-state rather than
the experimentally observed FM ground-state. This reversal of the
ground-state from FM to AFM exists for all $\beta$-values used to
determine the minimum energy (see Fig.~\ref{fig:tevsbeta}).
 
%figure14 Total energy vs Beta computation results

Other, more complex types of AFM arrangements ($C$-type, $G$-type,
etc.) can be constructed by generating larger supercells. However, if
these AFM cells were to generate lower energy configurations, as
compared to the $A$-type configuration that we used in the above
calculations, an increased (more negative) $\Delta E_{FM-AFM}$ would
result enhancing the difference with experiment. Instead, we focus
here on the $A$-type supercell exploring possible explanations for the
differences between the simulations, which yield an AFM ground-state,
and the experimentally observed FM low temperature state.

Experimentally the ground-state of Fe$_3$Ga$_4$ is FM with
antiferromagnetism accessed above 68 K in low magnetic
fields. However, we observed that $T_1$ can be somewhat lower for
several crystals prior to annealing which tends to sharpen the
transition and bring $T_1$ to 68 K.  Presumably the annealing reduces
disorder and residual stress in the crystals.  Thus, we have
considered the effect of disorder on the magnetic ground state
properties in our simulations. In Fe$_3$Ga$_4$, a simple atom counting
yields only $43\%$ Fe. The simulations described above show that a
small magnetic moment on the Ga sites is primarily due to the
interaction with the surrounding Fe. As Supplemental Materials Table~4
makes clear, the Ga nearest-neighbors are predominately Fe atoms. This
suggests that Fe$_3$Ga$_4$ has a strong ferromagnetic polarization
associated with Fe despite the larger concentration of Ga in the
system. The observation that the magnetic transitions are
substantially affected by a relatively low temperature anneal
indicates that there might be a small number of point defects in the
system, most likely antisite defects where a small number of Fe atoms
replace Ga. The idea is that a relatively small density of antisite
defects would thereby produce small FM Fe-clusters producing a larger
net polarization that potentially could lower the FM total energy
below that of the AFM phase.

To explore this possibility, FM and AFM simulations have been
performed on the $1\times 1\times 2$ supercells where two Ga atoms
have been replaced by two Fe atoms to mimic antisite defects.  For
this particular cell, an even number of antisites are necessary to
allow for the simulation of an AFM state. Simulations were performed
at the experimental $\beta$-value of 106.3$^\circ$ with the
translation vectors and the atomic position fully relaxed.  These
simulations produce a FM ground-state that is 1.24 eV per unit cell
lower in energy than the AFM state.  This substantial change is not
unreasonable given the rather large $\sim 4.8\%$ concentration of Fe
antisites.  To reduce the antisite density to levels closer to those
more likely to be present in the samples, we performed preliminary
fully relaxed calculations using a $4\times 4\times 4$ Monkhorst-Pack
$k$-space integration mesh on a larger $2\times 2\times 2$
Fe$_3$Ga$_4$  supercell where 2 Ga atoms are replaced by 2 Fe atoms
corresponding to a $\sim 1.2\%$ concentration of  Fe antisites. These
calculations yield a FM state that is $\sim 0.66$ eV/supercell lower
in energy than the AFM state, a significant drop in the energy
difference. A simple interpolation of the energy difference between
the two states based on the results of these simulations yields a
rough estimate of the minimum concentration of antisite defects
necessary to yield a FM ground-state of $\sim 0.1\%$ well below the
resolution of our X-ray diffraction measurements.

\section{Discussion and Summary}
We have presented an investigation of the magnetic, thermodynamic, and
charge carrier properties of Fe$_3$Ga$_4$ crystals to explore the
sensitivity of its magnetic states to temperature and magnetic
field. The measurements have not only established a moderate magnetic
anisotropy in this itinerant metamagnet, but have confirmed the main
features of the temperature-field magnetic phase diagram discovered in
polycrystalline samples. In addition, we have identified the specific
heat signal associated with the transition from the FM-to-AFM-like
state near 68 K as well as the larger, more diffuse, magnetic
contribution to the specific heat between room temperature and 400
K. The electrical transport is interesting because of the sharp change
in $\rho$ near $T_2$ which is not accompanied by equally discontinuous
changes in either $\chi$ or $C_P$. At the same time, the transition at
$T_1$ appears to be first order showing hysteresis upon warming and
cooling along with substantial changes in both $R_0$ and $R_S$, yet we
observe no discernible discontinuity in $\rho$. Furthermore, we
observe a significant $\rho_{THE}$ in the intermediate temperature
range between $T_1$ and $T_2$ suggesting a non-coplanar magnetic
state.

There are several classes of itinerant metamagnetic materials where
two magnetic phases, usually ferromagnetic and antiferromagnetic, are
close in energy and where temperature and magnetic fields can tip the
balance in favor of a ferromagnetic or field polarized PM state. These
instabilities can be accessed in a number of systems by tuning their
composition via chemical substitution between end members having
differing magnetic ground states. Examples include classic binary
compounds, Fe$_{1-x}$Rh$_x$\cite{fallot,shiraneferh}, layered
materials, Hf$_{1-x}$Ta$_x$Fe$_2$\cite{nishihara}, and shape memory
alloys, Ni$_2$Mn$X$ ($X=$ In, Sn, and Sb)\cite{sharma}.  However, this
tuning is not always necessary. Materials such as CoMnSi\cite{barcza}
and Mn$_3$GaC\cite{bouchaud}, are examples of stoichiometric compounds
that have similar metamagnetic properties and temperature dependent
magnetic states. The transitions between ferromagnetism and
antiferromagnetism are often accompanied by either a symmetry changing
structural phase transition or an abrupt change in unit cell
volume. In either case, large changes to the electronic density of
states at the Fermi energy are often apparent.

For the case of Fe$_3$Ga$_4$ the Hall effect data reveal changes
indicative of variations in the electronic structure near $T_1$ and
$T_2$. The jump in $R_0$, as well as the sign change of the AHE,
signaling a change to the reciprocal space Berry curvature, are
evidence of an abrupt variation in the electronic structure in
proximity to the Fermi energy. However, the very small specific heat
anomaly associated with the transition at $T_1$ is not consistent with
a strong magneto-elastic coupling as would occur if the magnetic phase
transition were accompanied by a change in the crystal structure.
Instead, we speculate that the changes to the electronic structure at
$T_1$ and $T_2$ are caused by entering and exiting a SDW phase. We
point out that the sharp rise in $\rho$ along with the reduction in
the carrier density suggested by $R_0$ at $T_2$ indicate a partial
Fermi surface gapping as would occur in a spin density wave state.
However, we have found no obvious nesting in the complex Fermi surface
that the simulations produce. In addition, the lack of a discontinuity
in $\rho(T)$ at $T_1$ where the system reenters the FM phase with
cooling, and the difference in sign of $R_S$ below $T_1$ and above
$T_2$, suggests that the sections of the Fermi surface gapped in the
purported SDW phase may not be completely recovered in the ground
state despite the changes we observe in $R_0$ near $T_1$.

Setting aside the character of the AFM phase, there remains a question
as to the cause of the close competition of FM and AFM phases in this
material. The complexity of the crystal structure which includes four
crystallographically distinct Fe-sites each with a different magnetic
moment is sure to play an important role in producing this
competition.  This idea is supported by our electronic structure
calculations which predict that an AFM ground state is slightly
favored over a FM one, but that a small density of antisite defects
can stabilize the FM phase. The sensitivity to synthesis technique and
annealing history of the samples that we observed would be explained
as a result of the crystalline disorder tipping the balance in favor
of the FM state. Thermal expansion may also play a role in driving the
system toward an AFM state as pressure and Al substitution for Ga tend
to stabilize the FM state to higher temperature\cite{duijnthesis}.
Thus, the AFM state may very well be the preferred phase at larger
lattice constants. However, because there are no indications of strong
magneto-elastic coupling in the specific heat of Fe$_3$Ga$_4$,
mechanisms involving a coupling of the magnetic degrees of freedom to
the lattice degrees of freedom are unlikely.

We have pointed out that Fe$_3$Ga$_4$ is unusual in that $H_{mm}$ {\it
increases} with $T$ when $H$ is parallel to the crystallographic
$c$-axis within the AFM state\cite{kawamiya1,kawamiya2}. This behavior
has also been reported in (Hf$_{1-x}$Ta$_x$)Fe$_2$\cite{nishihara} and
EuRh$_2$Si$_2$\cite{seiro} so that these may provide a useful point of
comparison. (Hf$_{1-x}$Ta$_x$)Fe$_2$\cite{wada,rechenberg} is unusual
in that the metamagnetism is thought to be associated with magnetic
frustration in the AFM phase. Here there are two distinct
crystallographic Fe sites, both of high symmetry. The $2a$ site is
also a point of inversion symmetry for this lattice. As a result, the
internal field is naturally canceled at the 2a site leading to
paramagnetic, highly fluctuating, magnetic moments persisting at
temperatures well below the Neel temperature. Thus, there is a
significant magnetic entropy associated with the AFM state above that
of the FM state tending to favor ferromagnetism at low $T$. In
contrast, the Fe$_3$Ga$_4$ crystal structure is not obviously layered
and frustration is highly unlikely\cite{kobeissi} so that a similar
mechanism is unlikely to play a role. In EuRh$_2$Si$_2$the low-$T$
phase is thought to be a spin spiral which transitions to fan-like
structure for fields along the easy plane\cite{seiro}. Such a model
may be possible for Fe$_3$Ga$_4$ given the layered planar structure
noted above. However, the nature of the anisotropy and the
discontinuous changes to $M(H)$ that we measure, see
Fig.~\ref{fig:m2}, make this explanation unlikely.

%The measurement of thermal expansion of Fe$_3$Ga$_4$
%by Duijn {\it et al.}\cite{duijnthesis} is instructive. These authors have
%measured the powder X-ray diffraction pattern from 5 to 295 K and
%report only a $0.7$\% increase in the the unit cell volume.  However,
%their data indicate an anisotropic thermal expansion as the c-axis
%expands at twice the rate of the a-axis over that temperature
%range. They also report only a very small anomaly at $T_1$
%corresponding to a change in volume of only 0.02\%. This is consistent
%with our specific heat data indicating only a small coupling of the
%lattice and magnetic properties associated with this
%transition. Unfortunately, their data do not extend above room
%temperature so that changes at $T_2$ and $T_3$ were not
%accessed. However, the anisotropic thermal expansion apparent in their
%data may be an important factor determining which magnetic phase is
%realized.  The structural complexity would thus be responsible for the
%fragility of the antiferromagnetic state which appears to be preferred
%at larger lattice constants.
  
With the lack of crystal symmetry produced frustration, and little
evidence for a magneto-elastic coupling, we are left to consider the
role of the itinerant electrons and the possibility of a SDW phase
along with the competing interactions caused by the four inequivalent Fe
sites within the unit cell as the cause of the rich magnetic behavior
we observe. It may be that Fe$_3$Ga$_4$ can be considered to be
intermediate between less complicated structural materials that have
little competition between possible magnetic ground states, and those
materials with enormous unit cells that produce spin-glass like
behavior without significant disorder or obvious magnetic frustration
present\cite{schmitt}. The drivers of such unusual behavior in
Fe$_3$Ga$_4$ are not obvious so that measurements of the magnetic
structure are clearly needed to make progress in understanding the
mechanisms for this unusual magnetic system.

\section{Acknowledgments}
Support from the NSF is acknowledged by JFD through DMR1206763, DPY
through DMR1306392, and JYC through DMR1358975. Support from the DOE
is acknowledged by PWA through DE-FG02-07ER46420 and JYC through
DE-FG02-08ER46528. This material is based upon work supported by the
National Science Foundation under the NSF EPSCoR Cooperative Agreement
No. EPS-1003897 with additional support from the Louisiana Board of
Regents.

\end{document}